\def\update{\Large }

\documentclass{mn2e}
\usepackage{graphicx}
\usepackage{epsfig}
\usepackage{float}

\newcommand{\BC}{\textsc{bc}}
\newcommand{\AD}{\textsc{ad}}

\newcommand{\gae}
{\lower 2pt \hbox{$\, \buildrel {\scriptstyle >}\over {\scriptstyle \sim}\,$}}
\newcommand{\lae}
{\lower 2pt \hbox{$\, \buildrel {\scriptstyle <}\over {\scriptstyle \sim}\,$}}

\newcommand{\0}{\phantom{0}}

\newcommand{\js}{2-1J$+$2S}

\begin{document}

\title[Three-body resonance in meteoroid streams]
{Three-body resonance in meteoroid streams}

\author[A. Sekhar, D. J. Asher, J. Vaubaillon]
{A. Sekhar$^{1,2,3*}$, D. J. Asher$^2$, J. Vaubaillon$^4$\\
$^1$Centre for Earth Evolution and Dynamics, Faculty of Mathematics and Natural Sciences, University of Oslo, Blindern N-0315, Norway\\
 $^2$Armagh Observatory, College Hill, Armagh BT61\ 9DG, United Kingdom\\
 $^3$Queen's University of Belfast, University Road, Belfast BT7 1NN, United Kingdom\\
 $^4$IMCCE, Observatoire de Paris, 77 Avenue Denfert Rochereau, F-75014 Paris, France\\
 $^*$E-mail: aswin.sekhar@geo.uio.no, asw@arm.ac.uk \\ }

\date{{\bf Accepted}: 2016 May 5 ; {\bf Received}: 2016 May 5 ; {\bf In Original Form}: 2015 May 23 ; {\bf MNRAS} \update}

\maketitle

\begin{abstract}
Mean-motion resonances play an important role in the evolution of various
meteoroid streams. Previous works have studied the effects of two-body
resonances in different comets and streams. These already established
two-body
resonances were mainly induced either by Jovian or Saturnian effects but
not both at the same time. Some of these resonances have led to spectacular
meteor outbursts and storms in the past. In this work, we find a new
resonance mechanism involving three
bodies -- i.e.\ meteoroid particle, Jupiter and
Saturn, in the Perseid meteoroid
stream. Long-term three-body resonances are not very common in real small
bodies in our solar system although they can mathematically exist at many
resonant sweet spots in an abstract sense in any dynamical system. This
particular resonance combination in the Perseid stream is such that it is
close to the ratio of 1:4:10 if the orbital periods of Perseid particle,
Saturn and Jupiter are considered respectively. These resonant Perseid
meteoroids stay resonant for typically about 2 kyr. Highly
compact dust trails due to this unique resonance phenomenon are present in
our simulations. Some past and future years are presented where three-body
resonant meteoroids of different sizes (or subject to different radiation
pressures) are computed to come near the Earth. This is the
first theoretical example of an active and stable
three-body resonance mechanism in the realm of meteoroid streams.

\end{abstract}


\section{Introduction} 

The fundamental concept of mean-motion resonances (MMR) in orbital dynamics
has been widely explored and studied by different authors. However,
correlating real examples in solar-system bodies with theoretical aspects of
MMR in celestial mechanics has been a great challenge ever since. To put
things into perspective in the context of this work, it is widely known that
influences of Jovian MMR in different comets and meteoroid streams (Chambers
1995; Asher \&
Emel'yanenko 2002; Jenniskens 2006; Ryabova 2006; Vaubaillon, Lamy \& Jorda
2006; Soja et al.\ 2011; Sekhar \& Asher 2014) can have a significant impact
on their long-term evolution. They play a big role in determining the
geometry and evolution of sub-structures in the meteoroid streams. Some of
these Jovian MMR have been directly correlated (Asher \& Clube 1993;
Jenniskens et al.\ 1998; Arlt et al.\ 1999; Asher, Bailey \& Emel'yanenko
1999; McNaught \& Asher 1999; Brown \& Arlt 2000; Rendtel 2007; Sato \&
Watanabe 2007; Christou, Vaubaillon \& Withers 2008; Sekhar \& Asher 2014)
with well-observed meteor outbursts and storms in the past. Such comparisons
and future predictions are one of the great applications of orbital studies
related to meteor showers. The effects of MMR due to other planets in various
meteoroid streams have not been studied at such length like Jovian  MMR. 
Nevertheless there are previous works exploring them in some detail
especially the ones related to Saturnian MMR (Brown 1999; Sekhar \& Asher
2013) and Uranian MMR (Williams 1997; Brown 1999) in meteoroid streams.

All these examples mentioned above are 2-body MMR. In these cases, the
individual meteoroid particle gets locked in resonance due to periodic
effects from one massive body (i.e.\ usually a planet). Hence this resonance
mechanism involves total of 2 bodies revolving around the central body (i.e.\
the Sun). There are plenty of examples, not just meteoroid stream particles,
of small bodies getting trapped in 2-body MMR in the solar system. However,
real examples of 3-body MMR in the solar system have been rarer; the first
well-known example was the Laplacian resonance in the Galilean satellite
system (Laplace 1799) involving Ganymede, Europa and Io exhibiting 1:2:4 MMR
respectively. The Laplacian relation has been studied at some length in the
past (see Murray \& Dermott 1999, section 8.16). After this well-known
discovery, a further example of natural satellites getting locked in 3-body
MMR, namely the Uranian satellites Miranda, Ariel and Umbriel, was apparently
identified but ceased to be a real 3-body MMR (Murray \& Dermott 1999, section 8.16)
after better observations which eventually led to improved orbital elements.
Later work (Nesvorn\'y \& Morbidelli 1998; Morbidelli \& Nesvorn\'y 1999)
showed that 3-body MMR plays an important role in the long-term evolution of
the asteroid belt. Subsequently Smirnov \& Shevchenko (2013) found that there
are a large number of asteroids trapped in 3-body MMR involving both Jupiter
and Saturn. Most recently Gallardo (2014) mapped an atlas of resonant
locations feasible for 3-body MMR in the solar system. 

Although the topic of 3-body MMR led to studies in the case of natural  
satellites and asteroids in the last few decades, none seems to have explored 
or made a systematic search for them in the realm of comets and meteoroid
streams. In this work, we investigate this problem and show conclusively that 3-body MMR, involving Jupiter, Saturn and a meteoroid, can occur in
the Perseid stream. Further calculations looking into the evolution of 3-body MMR meteoroids
show that meteor outbursts or storms on
Earth due to this unique resonance mechanism can be correlated in the past and predicted for the future. Perseids are one of the
most prolific annual showers (Rendtel 2014) and are well understood to have
originated (Jenniskens at al.\ 1998; Jenniskens 2006) from 109P/Swift-Tuttle
which is known to be an active comet. 

\section{Separation of 3-Body and 2-Body Resonances}
\label{sigma}
In Sections \ref{sigma}, \ref{geom}, \ref{resnonres} the MERCURY 
package (Chambers 1999) incorporating the RADAU algorithm (Everhart
1985) is used for simulating orbits in gravitational N-body integrations (Sun + 8 planets). Osculating elements for planets were from JPL Horizons 
(Giorgini et al.\ 1996). For 109P they were from Marsden \& Williams (2008); 
cometary non-gravitational parameters are unavailable as Marsden et al.\
(1993) concluded that 109P's observations back to 69 \BC\ can be fitted
gravitationally. Therefore effects like non-gravitational forces, Yarkovsky
and YORP are not included for the parent body 109P/Swift-Tuttle.  

\begin{table}
\centering
\caption{Resonant configurations, order of resonances, mean resonance
locations and approximate strengths in decreasing order of
strength (explained in Gallardo 2014) for different 3-body MMR
for nominal Perseid orbital elements.}
\label{MMRlist}
\begin{tabular}{@{}rcccc@{}}
\hline
MMR &Order of & Mean Resonance & Strength     \\
  & Resonance   q         & Location $a_{n}$ (au) & $\times 10^{-2}$     \\ 
  &                                     &                                  &  \\
\hline
2-1J+2S    & \03        & 24.41    & 7.021  \\
6-1J+1S    & \06        &24.19    & 1.567 \\
8-2J+3S   & \09        &24.24      & 0.695\\
14-1J-6S    & \09        & 24.03   & 0.218 \\
10+1J-5S    & \06        &24.01    & 0.201\\
2+1J-3S   & \00        &23.75     & 0.168 \\
\hline
\end{tabular}\\
\end{table}

In this work we focus on true 3-body resonances involving Perseid particles with both
Jupiter and Saturn. Previous works (Jenniskens et al.\ 1998;
Emel'yanenko 2001) have shown the existence of 1:10, 1:11 and 1:12 Jovian MMR in Perseids. Our test simulations independently indicated the possibility for 1:4 Saturnian MMR (not shown here) as well. 

This inspired us to check for a possibility of 3-body resonance with triple
ratio 1:4S:10J involving both these planets. Saturn and Jupiter are themselves near a 2:5 commensurability
(but never in exact resonance) called the `Great Inequality' (Murray \& Dermott 1999, page 10). Counterintuitively, for 3-body MMR to exist, it is 
not necessary that two individual pairs are resonant with
each other. 

Gallardo (2014) has looked into the abstract cases of 3-body resonances in 
the solar system and mapped locations (in terms of the `nominal resonance
location' $a_{n}$) favourable for such resonances. Circular coplanar orbits
are assumed for two planets and the resonance locations can be calculated for
test particles with any value of eccentricity $e$ and inclination $i$. This
technique gives an estimation of strengths of resonances and is very useful
to compare between different possible 3-body MMRs and distinguish between the
weakest and strongest MMRs. Hence we use this method here, to find the
strongest 3-body MMR resonance candidate locations in the Perseid stream.

In terms of mean motions (cf.\ Eqn (\ref{eqn-sigma}) below) the
configuration closest to this 1:4S:10J resonance location is 2-1J$+$2S, for which the semi-major axis $a$ takes the value
$a_{n}$=24.41 au. Using the code developed by Gallardo (2014), available at
\texttt{www.fisica.edu.uy/$\sim$gallardo/atlas}, we recalculate all the
possible lower-order resonances: order of resonance q$\leq$10 for $a$=23 to
26 au with degree p $\leq$ 20 for the nominal $e$=0.95, $i$=113\degr,
$\omega$=150\degr\ of Perseids, these $e$, $i$, $\omega$ being taken from
IAU-MDC (Meteor Data Center).
The result yielded 14 MMRs with strengths ranging from 10$^{-6}$ to 10$^{-2}$
out of which two MMRs have significantly higher strengths by an order of
magnitude compared to other configurations (Table \ref{MMRlist} shows the
first six strong candidates, amongst all the 14 MMRs displayed by the code,
in descending order of strength). The first and sixth highest strength MMRs
(which are discussed in this work) relevant in the case of Perseids are at
nominal resonance locations $a_n$=24.41 au (corresponding resonant
configuration 2-1J+2S) and 23.75 au (corresponding resonant configuration
2+1J-3S) respectively.  For an approximate estimate of $a_n$, one
can take planetary mean motions considering each
individual planet to be on an unperturbed Kepler orbit, and apply Gallardo's
equation 2 using masses and mean semi-major axes of Jupiter and Saturn
(masses and mean $a$ values may be taken, for example, from Bretagnon 1982). 

As discussed in Gallardo's section 2 the actual location of the MMR depends 
on the precession of perihelia and on all the gravitational
effects of the planets. With $a_n$ = 24.41 we find ratios of orbital periods $P_{m} = 4.08
\times P_{S} = 10.16 \times P_{J}$ (cf.\ 1:4S:10J; in terms of \js\ note that 
$2 \times 1 - 1 \times 10.16  + 2 \times 4.08 = 0$). Here m, S and J correspond
to meteoroid particle, Saturn and Jupiter respectively. The osculating $a$ of
any resonant particle librates, approximately about $a_n$.

Table \ref{MMRloc} gives the resonance locations $a_{n}$ and order q for the
2-body and 3-body MMRs discussed in this work. The order of the 3-body MMR is
q=$\mid$$k_{0}+k_{1}+k_{2}$$\mid$ ($k_i$ as defined in Eqn \ref{eqn-sigma}
below). The order of an exterior 2-body MMR of configuration p:(p+q) is q. Typically the lower the order of a
resonance, the stronger is the resonance mechanism (if the same planet is
involved) for low eccentricities. However, for highly eccentric orbits (like the case discussed here) there could be several overlapping higher order resonances with different strengths as shown by Table \ref{MMRlist}. 

\begin{table}
\centering
\caption{Mean resonance location and order of resonances for the 3-body MMR
and nearest 2-body MMRs in Perseids. Although the technical 
nomenclature for this particular 3-body MMR in Perseids is \js and 2+1J-3S, the ratio of
orbital periods can be approximated to 1:4S:10J for Perseid particle, Saturn
and Jupiter respectively. P denotes the approximate interval 
(cf.\ Sekhar \& Asher 2013) until the next series of successive encounters of
the same resonant cloud with Earth (see Section \ref{geom} later), and
corresponds to idealized (without other planetary effects) orbital periods. 
}
\label{MMRloc}
\begin{tabular}{@{}rcccc@{}}
\hline
MMR &Order of & Mean Resonance  & P   \\
  & Resonance   q         & Location $a_{n}$ (au)     &(yr) \\ 
  &                                     &                                 &      \\
\hline
1:4S    & \03        & 24.03 &118     \\
1:10J    & \09        &24.14  &119   \\
 &       &  & \\
1:4S:10J   & \03        &24.41  &121  \\
\js   &       &  & \\
 &       &  & \\
1:4S:10J   & \00        &23.75  &116    \\
2+1J-3S   &       &  &    \\ 
\hline
\end{tabular}\\
\end{table} 

As expected, 3-body MMR is much more complicated (Yoder \& Peale 1981) than 2-body MMR and hence one has to be careful while analysing these resonances.
Hence it is critical that we do extensive tests to identify a 3-body MMR,
rigorously eliminating the strongest 2-body MMRs closest to the resonance
location connected with the particular 3-body MMR of our focus. 

Our method to confirm 3-body MMR in Perseids is to analyse the evolution of
critical angle (or resonant argument) $\sigma$ for given resonances.
For a 3-body MMR, this is defined by:
\begin{equation}
\sigma = k_{0}\lambda_{0}+k_{1}\lambda_{1}+k_{2}\lambda_{2}-(k_{0}+k_{1}+k_{2})\varpi_{0}
\label{eqn-sigma}
\end{equation}
where $k_{i}$ are integers, $\lambda_{i}$ are mean longitudes and $\varpi$ is
longitude of pericentre. Although there are different possible critical
angles, this particular one is the principal critical angle because it is
associated in the disturbing function with the term factorized with the
lowest-order power of $e$ (cf.\ Gallardo 2014).
Because Perseids are retrograde, we define $\varpi
= \Omega - \omega$ (Saha \& Tremaine 1993; Whipple \& Shelus 1993) and not
$\Omega + \omega$ ($\omega$ is argument of pericentre, $\Omega$ is longitude
of ascending node). The resonance configuration is
$k_{0}+k_{1}P_{1}+k_{2}P_{2}$ if the resonance involves planets $P_{1}$ and
$P_{2}$:  we follow this notation
throughout. In our work, the resonant particle is a Perseid and the planets
are Saturn and Jupiter. 

For \js, the critical angle
$\sigma=2\lambda_m-1\lambda_{J}+2\lambda_{S}-3\varpi_m$ is plotted versus
time. The key logic for verification is: if
$\sigma$ librates versus time and if there is a correlation in the time evolution of semi-major axis with the critical angle, resonance is confirmed. On the
other hand, if $\sigma$ circulates versus time, it rules out that configuration of resonance.

\begin{figure}
(a)\\[-\baselineskip]
\includegraphics[width=\columnwidth]{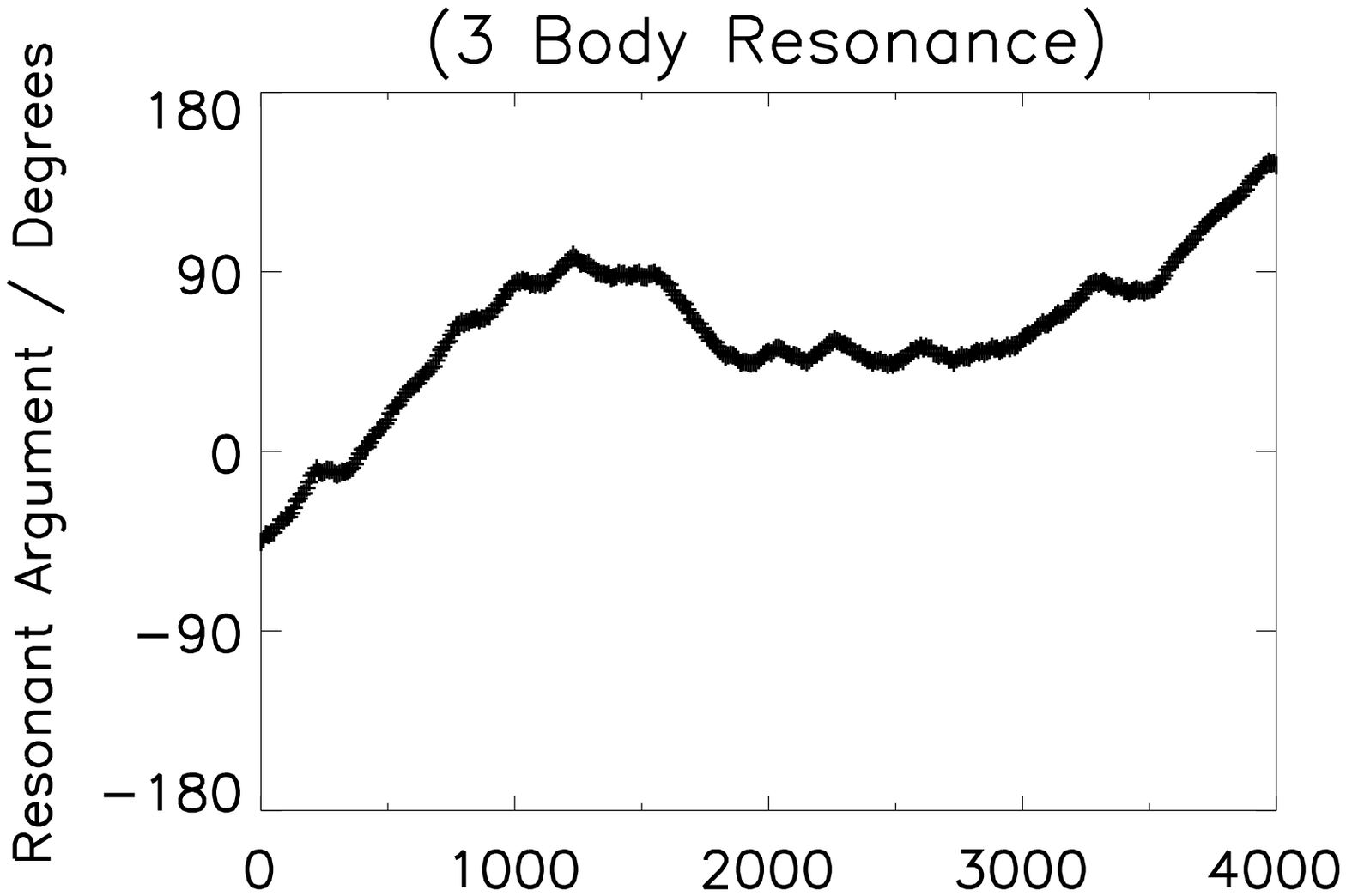}
\\[-5mm]
(b)\\[-\baselineskip]
\includegraphics[width=\columnwidth]{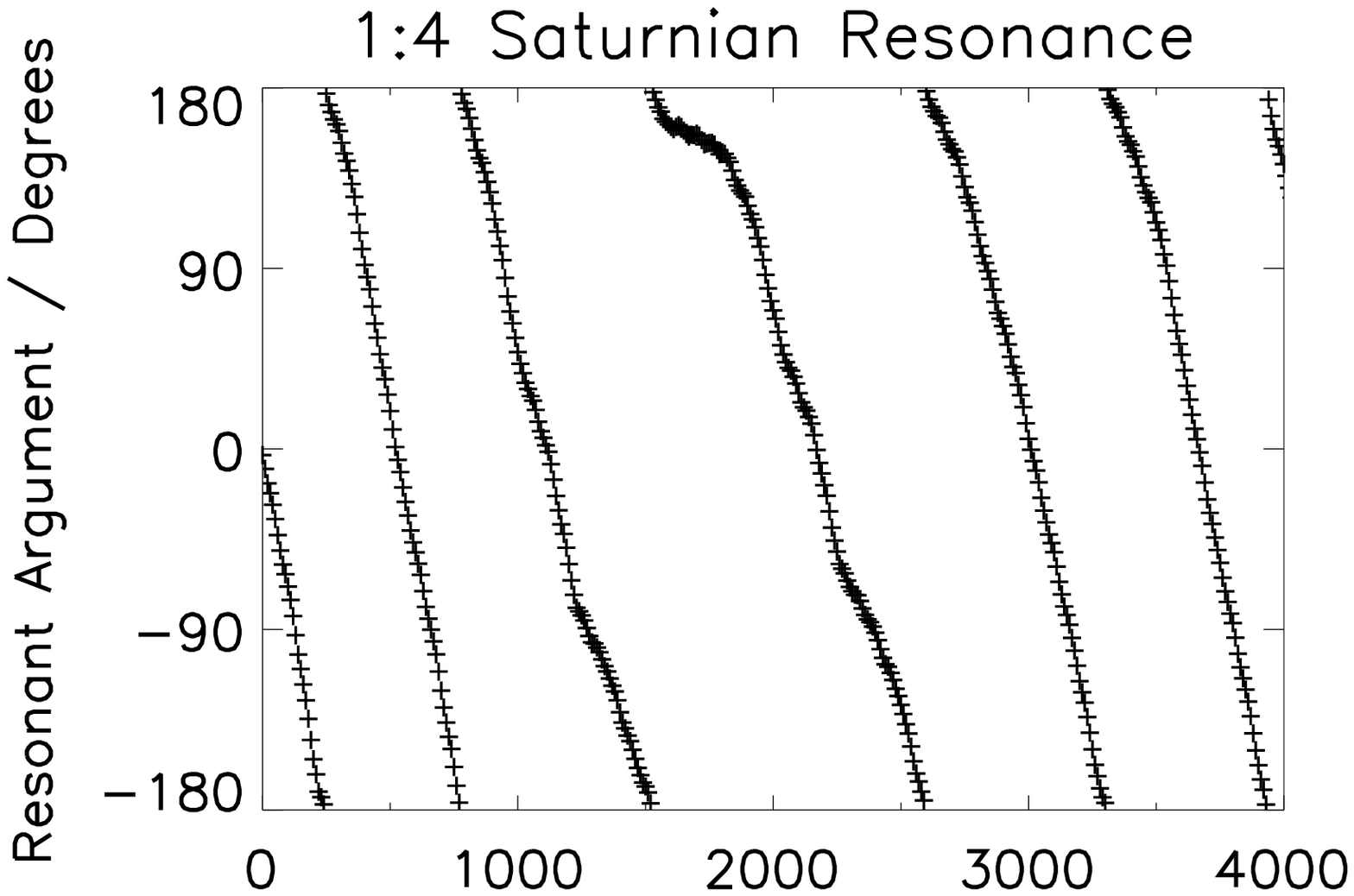}
\\[-5mm]
(c)\\[-\baselineskip]
\includegraphics[width=\columnwidth]{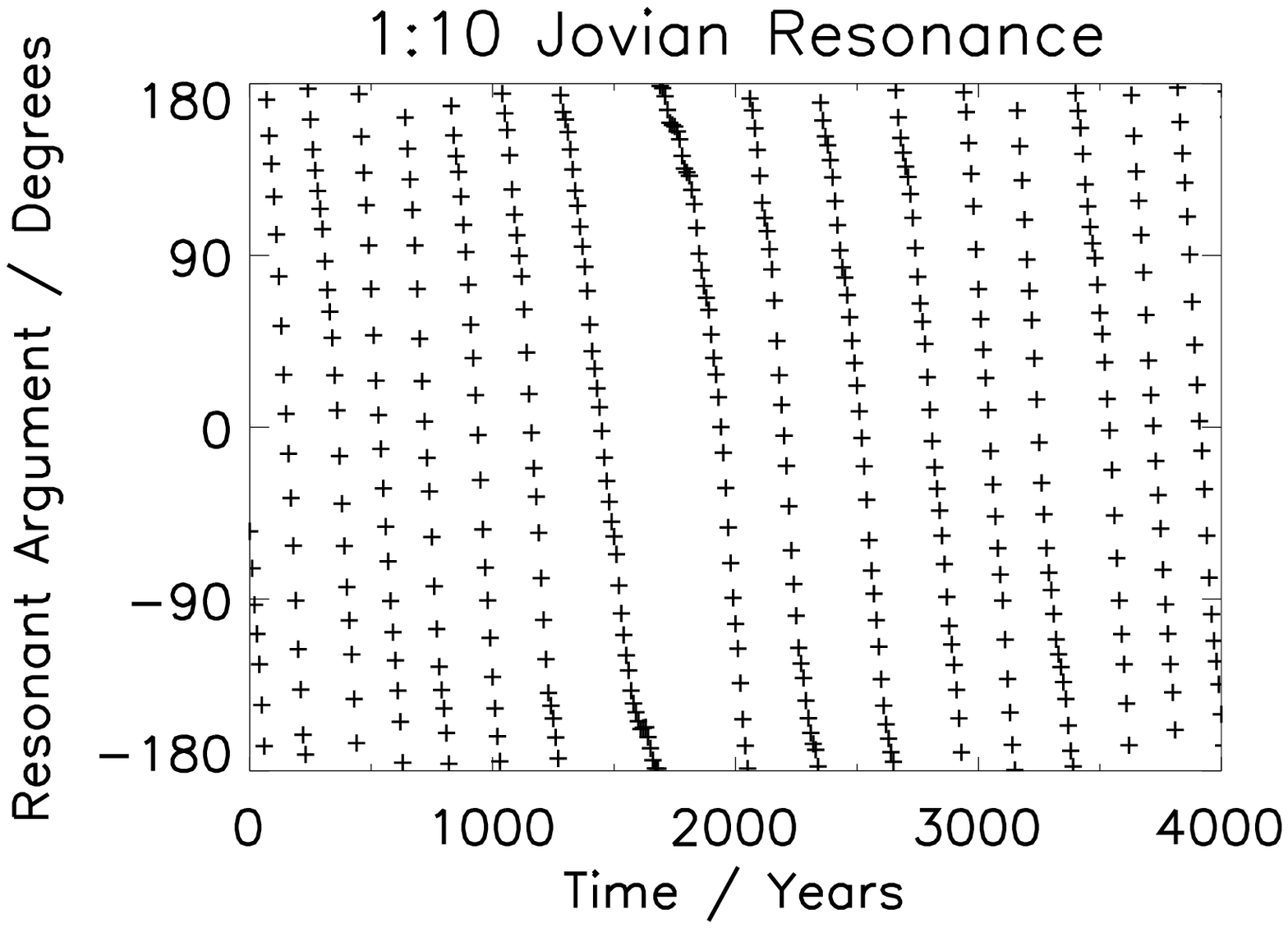}
\caption{(a) Libration of \js\ resonant argument for a Perseid
test particle (initial $a$=24.334 au), confirming presence of 3-body MMR 
involving Jupiter and Saturn simultaneously.  Outer circulation of (b) 1:10
Jovian (2-body MMR) and (c) 1:4 Saturnian (2-body MMR) resonant arguments for
the same particle during same time frame confirm absence of 1:10 Jovian
and 1:4 Saturnian separately. Furthermore absence of
resonance in (b) and (c) can be directly compared and confirmed with
well-defined
outer circulation trajectories shown in Figure \ref{PERtraj} (a) and
(b). Starting epoch (zero time) is JD 1696460.0 = 69 \BC\ August 27.5, the
oldest known return of 109P/Swift-Tuttle (Marsden \& Williams 2008).} 
\label{PERsigma}
\end{figure}

\begin{figure}
(a)\\[-\baselineskip]
\includegraphics[width=\columnwidth]{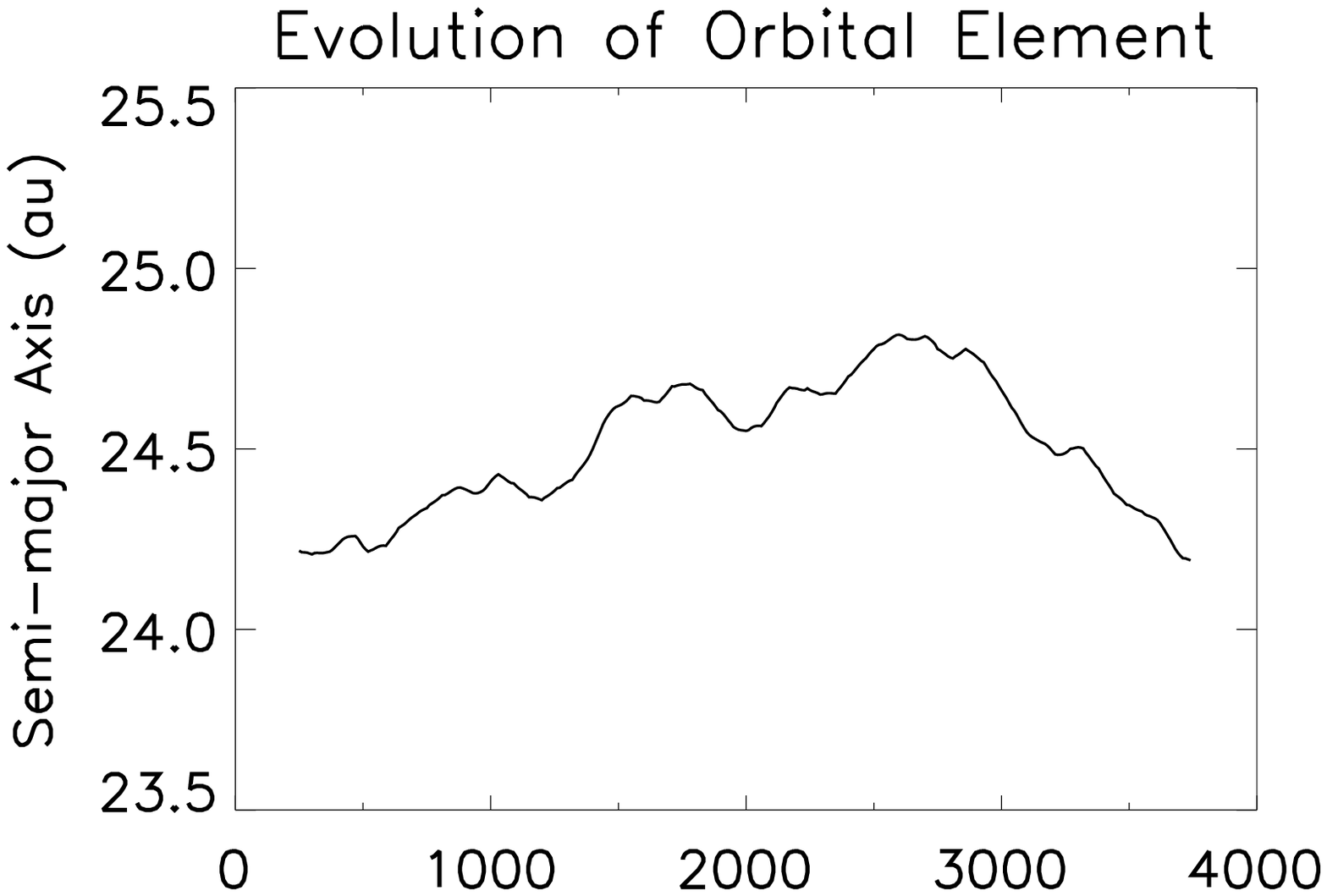}
\\[-5mm]
(b)\\[-\baselineskip]
\includegraphics[width=\columnwidth]{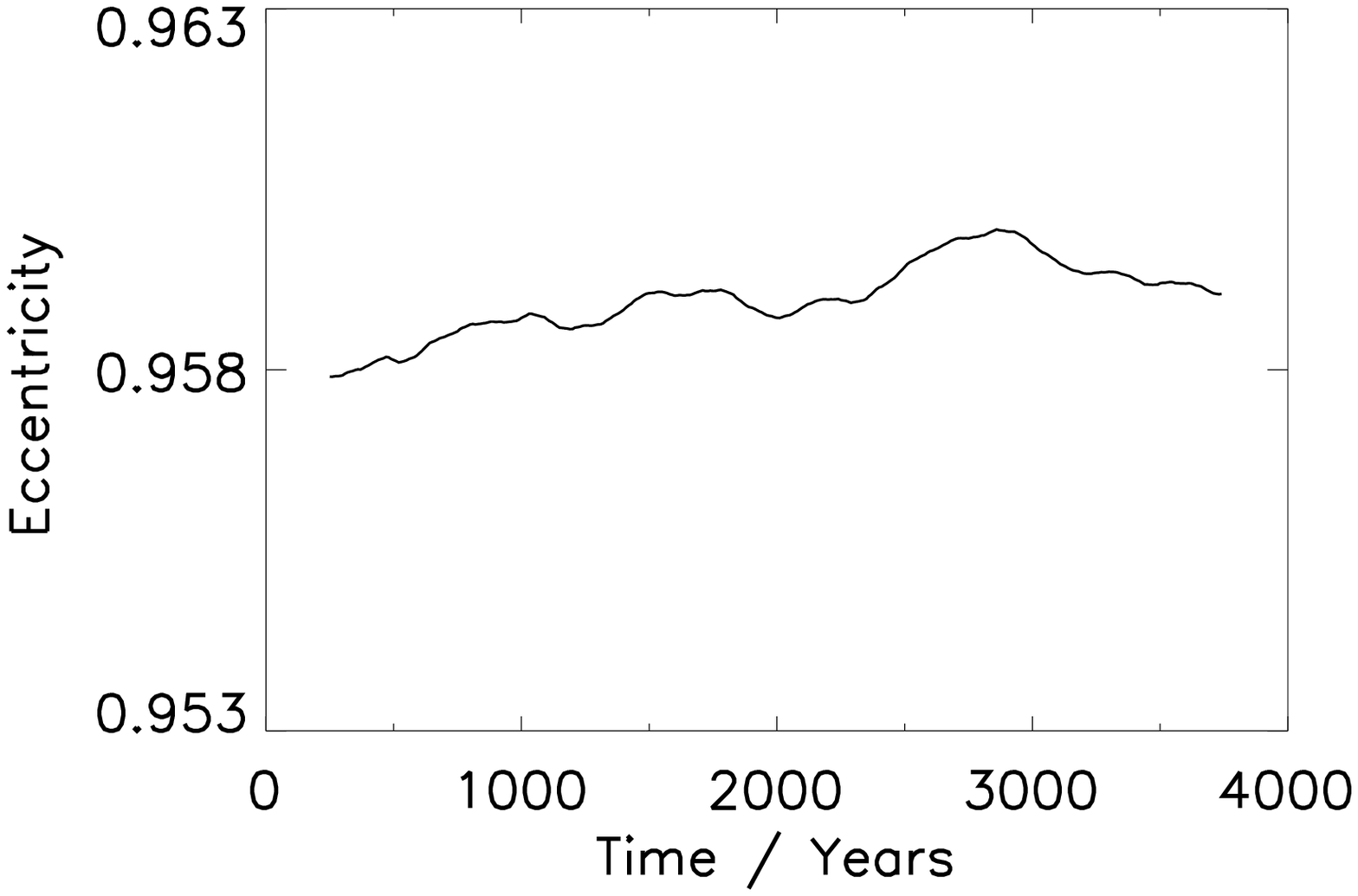}
\caption{(a) Libration in moving averaged (running window of 500 yr) $a$ matching with libration in \js\ resonant
argument for the same Perseid test particle as in Figure \ref{PERsigma},
indicating presence of 3-body resonance with Jupiter and Saturn. (b) Evolution of moving averaged (running window of 500 yr) $e$ for the same Perseid test particle as in Figure \ref{PERsigma}. }
\label{PERMMR}
\end{figure}

\begin{figure}
(a)\\[-\baselineskip]
\includegraphics[width=\columnwidth]{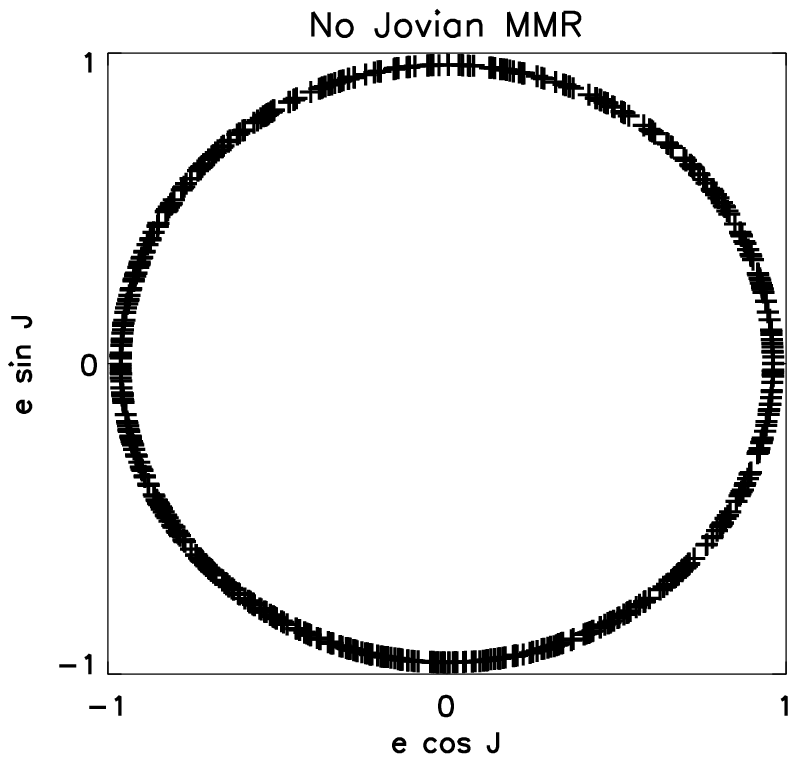}
\\
(b)\\[-\baselineskip]
\includegraphics[width=\columnwidth]{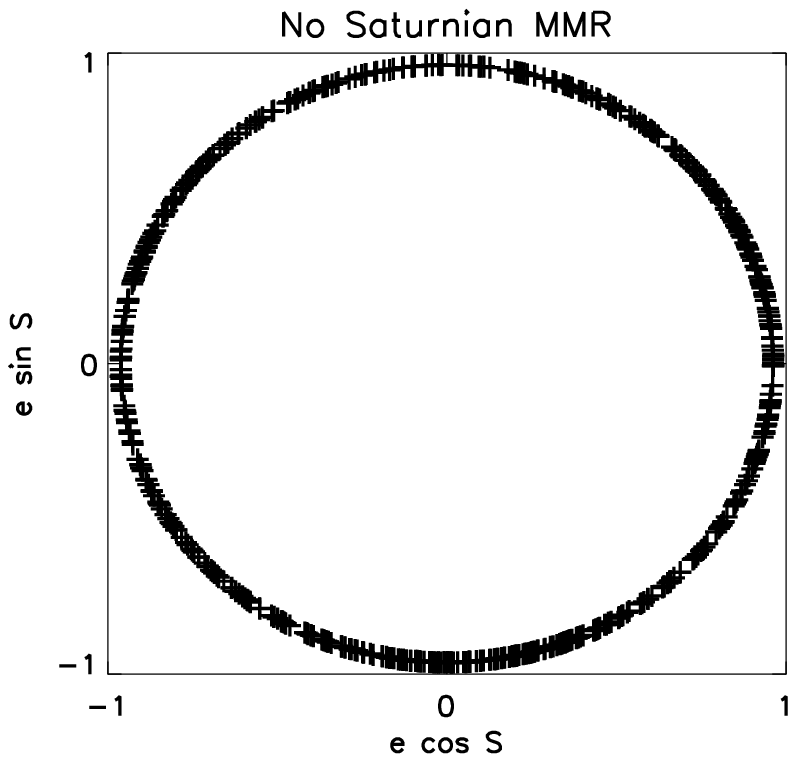}
\caption{ Trajectory phase space for (a) 1:10 Jovian (2-body MMR)
and (b) 1:4 Saturnian (2-body
MMR) cases for same particle during same time frame show outer circulation
and hence confirm absence of either of these resonances over same time frame
as shown in Figure \ref{PERsigma}.
In (b), J=$\sigma_J=\lambda_J-10\lambda_m+9\varpi_m$ and
in (c), S=$\sigma_S= \lambda_S-4\lambda_m+3\varpi_m$. }
\label{PERtraj}
\end{figure}

\begin{figure}
\includegraphics[width=\columnwidth]{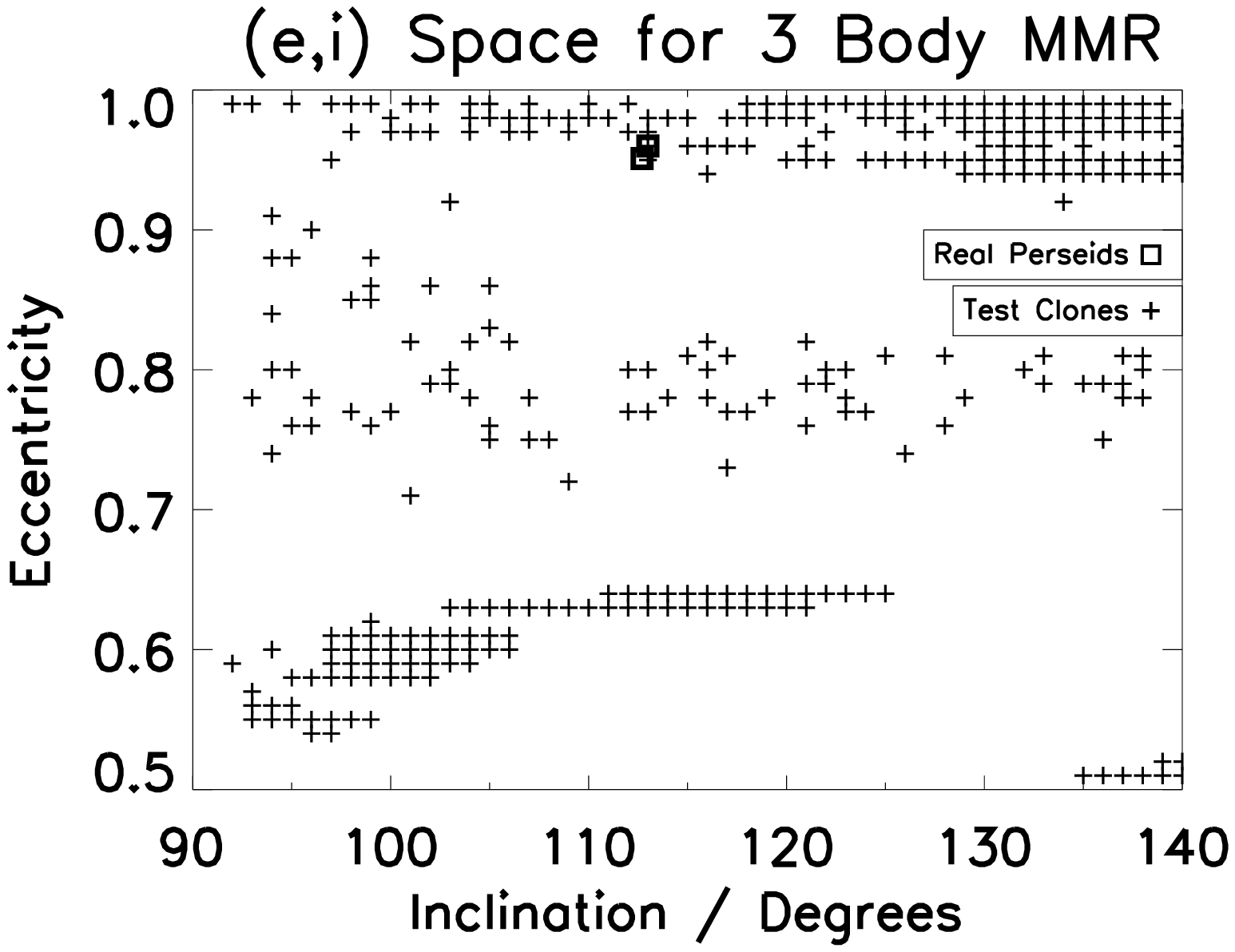}
\\[-3mm]
\caption{Stable regions of ($e$, $i$) space favourable for \js\ MMR in
Perseid-like orbits. The region of this phase space for real
Perseids ($e$, $i$ as given by IAU-MDC) is shown. Points
plotted are particles remaining \js\ resonant (3-body
MMR) for 4 kyr from 69 \BC\ perihelion passage.}  
\label{eispace}
\end{figure}

Nesvorn\'y \& Morbidelli (1999) presents an analytical theory for 3-body MMRs in the solar system for planar orbits and discusses the estimation of libration amplitudes and libration timescales. Nesvorn\'y \& Morbidelli (1998) has shown that the approach using this critical angle defined above works for orbits with different eccentricities. Hence it is logical to use this critical argument and analysis technique for the cases mentioned above. So the major task is to plot the relevant critical angles for this 3-body MMR
as functions of time and verify the nature of their evolution.  

Figure \ref{PERsigma}(a) shows the \js\ critical angle $\sigma$ librating
continuously for about 3 kyr. To check
the presence or absence of the 2-body MMRs (cf.\ Sekhar \& Asher 2013) we
plot resonant arguments
$\sigma_{S} = \lambda_S- 4\lambda_m+3\varpi_m$ for 1:4 Saturnian and
$\sigma_{J} = \lambda_J-10\lambda_m+9\varpi_m$ for 1:10 Jovian.
Figure \ref{PERsigma}(b) and (c) shows $\sigma_{J}$ and $\sigma_{S}$ clearly
undergoing outer circulation during the same time frame for the same particle.

Figure \ref{PERMMR}(a) and (b) shows the libration in semi-major axis and eccentricity
corresponding to the same time frame for the same particle shown in Figure
\ref{PERsigma}. Resonance can be confirmed if there exists a correlation between the librations in the critical angle and the semi-major axis of the particle;
this can be seen here. The phase space resonant
angle trajectories for 1:10 Jovian and 1:4 Saturnian MMRs were analysed (Murray \& Dermott 1999, section 8.9).
Figure \ref{PERtraj}(a) and (b) clearly show outer circulation for the same Perseid particle for
4 kyr. This definitely rules out the existence of both these individual 2-body MMRs during the same time frame for the same particle.  

The crucial point is to conclusively
establish the absence of 2-body MMR cases separately. It is well understood
that Jupiter and Saturn are not in 2-body MMR and hence we have not repeated
those calculations explicitly here. The absence of all these three individual
pairs undergoing 2-body MMRs do not pose any threat to 
the existence of 3-body MMR, as discussed in Greenberg (1975). 

Because our work deals with real comets and meteoroid particles, we did a
systematic study to verify the existence of the 3-body MMR \js\ in terms of
varying ($e$, $i$) orbital space,
comparing regions of ($e$, $i$) phase space both away from and
near the nominal elements of the Perseid stream. Figure \ref{eispace} shows the stable regions favouring \js,
for trapping particles in resonance. It can be seen that high eccentricity is
more favourable for MMRs while low eccentricity and near 90\degr\ inclination
combinations are least favourable for MMR. It is vital to confirm that the
3-body MMR phenomenon does not break down near the real Perseid orbital
element phase space and this was the primary motivation for these
tests.

\section{Extent of Resonant Zones}
\label{geom}

\begin{figure}
\includegraphics[width=\columnwidth]{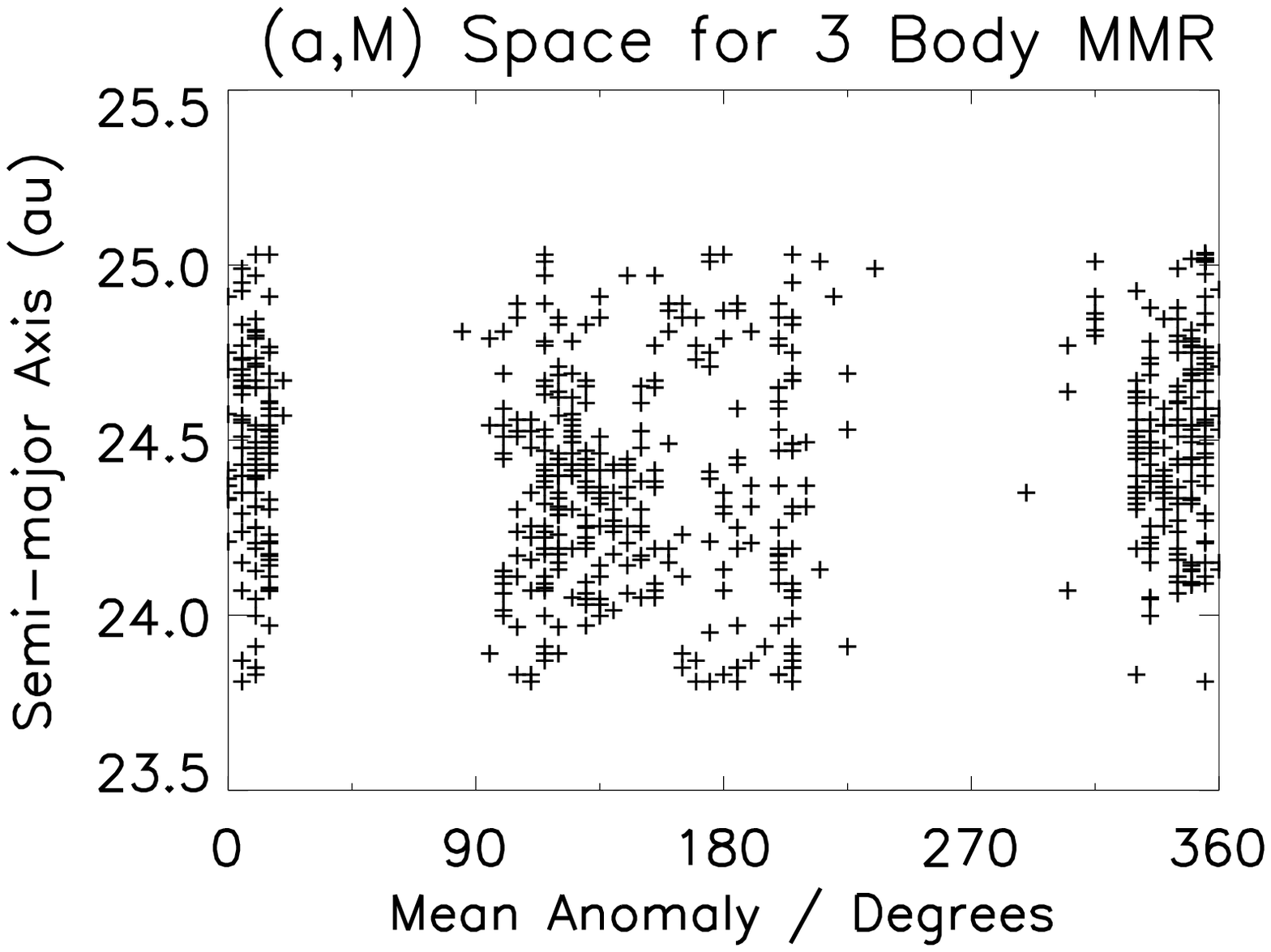}
\\[-3mm]
\caption{Resonant zones for \js\ MMR in Perseids in terms of initial $a$ vs
initial $M$. Points plotted are particles remaining \js\ resonant
(3-body MMR) for 4 kyr from 69 \BC\ perihelion passage.
}
\label{aM}
\end{figure}

Visualizing the effects of resonance on internal stream structure is
important to understand the overall collective behaviour of resonant
particles whose individual evolution with time was
discussed in Section \ref{sigma}. For this purpose we plot initial
semi-major axis $a$ and initial mean anomaly $M$ favouring the 3-body MMR
mechanism from the starting epoch of 109P/Swift-Tuttle's observed 69 \BC\
return.  Figure \ref{aM} shows the geometry of resonant zones for \js\
libration in the Perseid stream. Two resonant zones can be seen,
spanning two different ranges of $M$. These zones are like potential wells
connected with this 3-body MMR and particles trapped in these energy points
would librate along the boundary of the zone trajectories (in case of
high-amplitude
librations in critical angle $\sigma$) or librate very close to the
centre of zones (for extremely low-amplitude librations in $\sigma$). In a
fundamental sense, the motion of particles in resonant potential wells are
compared with the behaviour of a simple pendulum (section 8.6,
Murray \& Dermott 1999). If the particle reaches close to a
separatrix, then small effects from gravitational or other forces could
disturb the resonance and make the particle circulate (i.e.\ become
non-resonant) over time. 

The modus operandi of obtaining this plot involved generating 7200 particles
by varying initial $a$ from 23.43 to 25.41 au in steps of 0.02 au, and
initial $M$ from 0 to 360\degr\ in steps of 5\degr, keeping $q$, $i$,
$\omega$, $\Omega$ the same as parent body 109P ($q$ = perihelion distance).
Particles are integrated 4 kyr and only those trapped continuously in this
3-body MMR for 4 kyr are plotted here: for each particle, $\sigma$ is
evaluated for the entire 4 kyr in time steps of 10 yr and a histogram of
the resulting $\sigma$ values is created with bins of 10\degr.  If at least
one bin remains void, the particle is flagged as resonant (considering that
$\sigma$ oscillates around some libration centre in such cases). If all bins
are populated, the particle is assumed to have circulated through 360\degr\
and is flagged as non-resonant. Figure \ref{aM} shows
the locations of these resonant particles in $(a,M)$ space at the initial
69 \BC\ epoch.

The picture of resonant zones in this phase space 4 kyr later, or indeed at
any different time until 4 kyr, is similar to Fig.\ \ref{aM}: at any time
there are two zones spanning approximately the same total range in $M$ and
$a$. The central $M$ value of each zone moves progressively forward depending
on the orbital period of the MMR; we find it takes about 121 yr to complete
one 360\degr\ revolution in $M$.

The orbital period of the parent body during these millennia remains within a
few yr of 132 yr (Marsden et al.\ 1993 table III). Therefore over time the
comet drifts backwards through the two resonant zones and can populate both
of them with meteoroids, creating clouds of particles in the resonant zones.

P is defined as the time between successive encounters (see Table \ref{MMRloc})
of Earth with the same resonant cloud (Sekhar \& Asher 2014). Previous
work (Asher et al.\ 1999; Emel'yanenko 2001; Sekhar \& Asher 2013, 2014)
found P $\sim$ 71 yr for 1:6 Jovian, and P $\sim$ 88 yr for 1:3
Saturnian in the case of Orionids; P $\sim$ 33 yr for 5:14 Jovian, and P
$\sim$ 33 yr for 8:9 Saturnian in the case of Leonids.  

The trajectory of the comet inside the zones over time may not be uniform and
hence different parts of a zone get populated in different intensities and
shapes, leading to fine structures inside the zones.
These structures would vary greatly depending on ejection conditions (i.e.\
initial $a$ during each specific perihelion passage and whether this leads
to high- or low-amplitude libration in $\sigma$), subsequent time
evolution and some parts of zones having future close approaches with planets.
Predicting an exact intersection leading to a meteor outburst or storm on
Earth is directly dependent on whether the Earth traverses through one of
these clouds and intersects fine structures or not. Such an intersection would drastically
enhance the density of the
meteoroid population encountered by Earth or in other words, the Zenithal Hourly
Rate (ZHR) could significantly increase during such an event.

Because the gaps between resonant zones are significant
(Fig.\ \ref{aM}), it is unlikely that
Earth would intersect one of the zones on a regular basis every time. But
Earth missing these resonant zones does not imply zero meteor activity for
Perseids, pointing rather to normal activity without any additional boosting
of ZHR due to this particular 3-body resonance mechanism. Hence one should be
careful not to visualize Earth hitting these discrete clouds as the only
reason for substantial Perseid activity on Earth.  

Although this represents only a general picture of the potential wells
associated with this 3-body MMR in space, the simulations show that \js\
resonant meteoroids can lead to distinct resonant cloud evolutions clearly
for many kyr (2 kyr is typical). 
The range in $a$ spanned by the resonant cloud (Fig.\ \ref{aM}) is
equivalent to tangential ejection velocities in the range $\sim$ 4 to
29 m\,s$^{-1}$ (approximately 16 m\,s$^{-1}$ to populate the centre of the zone), directed backward of the comet's heliocentric motion, at the
69 \BC\ return ($a_\mathrm{comet}$=25.41 au) of 109P/Swift-Tuttle. These 
ejection velocities are realistic in cometary activity (Whipple 1951; Jones 
1995; Crifo \& Rodionov 1997): the Whipple formula gives 72 m\,s$^{-1}$ for a particle of diameter 1 mm
and Perseid density 2.25 g\,cm$^{-3}$ (table 2, Babadzhanov \& Kokhirova
2009) ejected from a comet nucleus of radius $\sim$13 km (Lamy et al.\ 2004)
at 109P's perihelion distance in 69 BC. As smaller velocities than this value
of 72 m\,s$^{-1}$ are sufficient to effectively populate this 3-body MMR,
there is more likelihood of bigger particles getting trapped into resonance,
in turn suggesting brighter meteors. Compared to many comets, 109P's
relatively large nucleus size favours higher velocity ejections
and therefore larger particles ejected at a given velocity.
The equivalent particle sizes that have the required range in ejection
velocities (again, considering the tangential component only) $\sim$ 4 to
29 m\,s$^{-1}$ are 5 cm to 0.6 cm respectively. The 1737 \AD\ return is similarly 
favourable in terms of lower ejection velocities, because $a_\mathrm{comet}$ is lower than at other observed 
returns, but the positioning in $M$ of the comet relative to the closer of 
the two resonant zones is not favourable as in 69 \BC\ to populate that
resonant zone effectively. All other observed returns require slightly higher
ejection velocities to populate resonant locations and indeed no other
observed perihelion return except that of 1862 \AD\ is favourable in terms of
the position of the comet to populate resonant zones of 3-body MMR locations
discussed in this work. This is the main reason why 
the 69 \BC\ return was chosen as the starting epoch here (further discussion
in Section \ref{resnonres}).

On an independent note, the numerical libration width in $a$ for this 3-body
MMR from our calculations is about 1.2 au which is in turn close to the
semi-analytical libration width of 1.4 au for 1:10 Jovian MMR in Perseids
(table 1, Emel'yanenko 2001). This gives greater confidence in our
results. Our past work gives a libration width of about 1.2 au (figure 3,
Sekhar \& Asher 2014) for 1:6 Jovian MMR in Orionids and 0.1 au (figure
2.11, Sekhar 2014) for 5:14 Jovian MMR in Leonids which were close to
libration widths calculated by Emel'yanenko (2001) of 1.0 au and 0.13 au
respectively. Small differences between analytical and numerical
libration widths are normal because the survival times, stability and close
encounters (with planets) of these particles depend on initial conditions and
duration of integrations.

\section{Resonant versus non-resonant particles}
\label{resnonres}

\begin{figure}
(a)\\[-\baselineskip]
\includegraphics[width=\columnwidth]{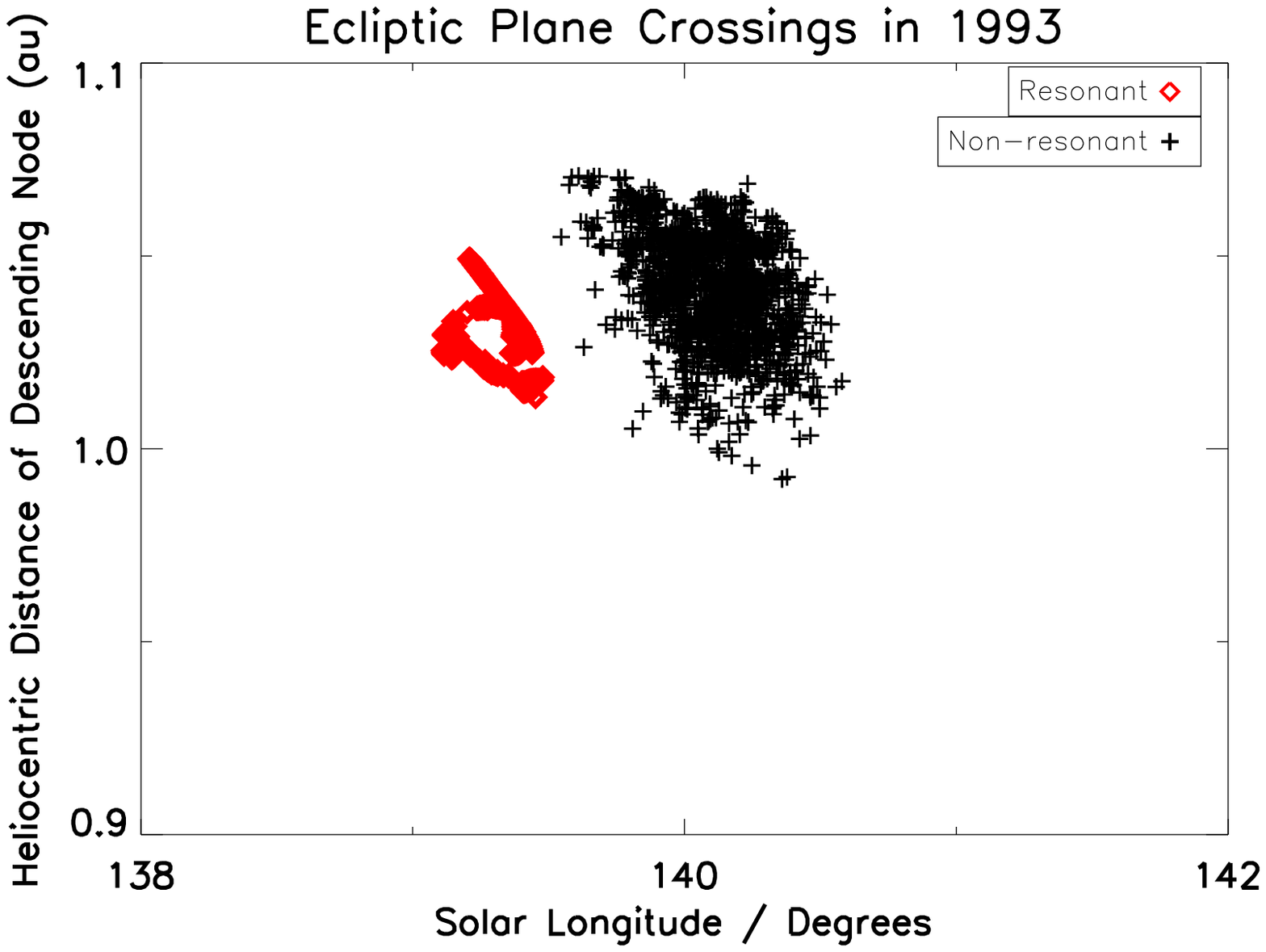}
\\[-3mm]
(b)\\[-\baselineskip]
\includegraphics[width=\columnwidth]{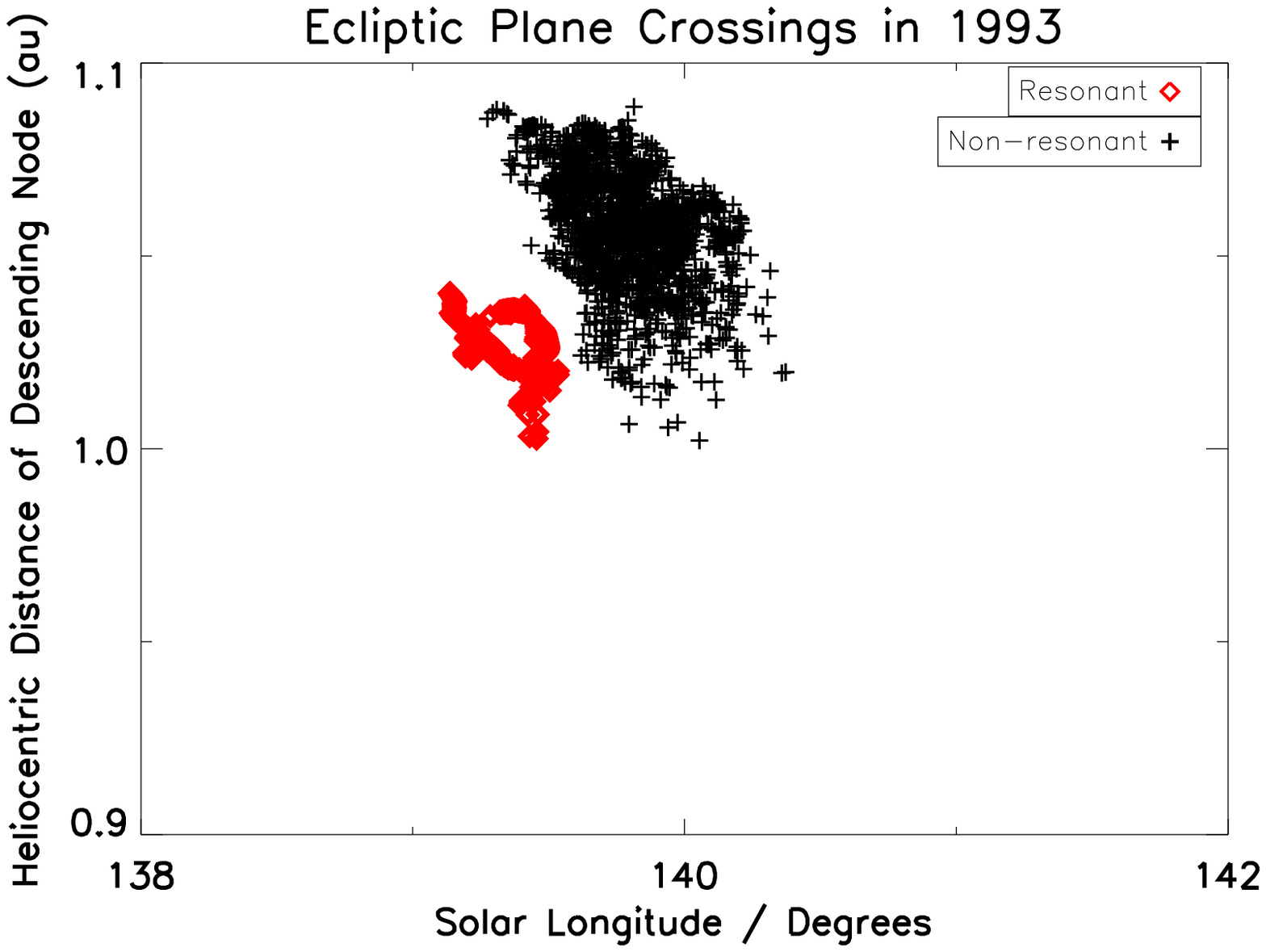}
(c)\\[-\baselineskip]
\includegraphics[width=\columnwidth]{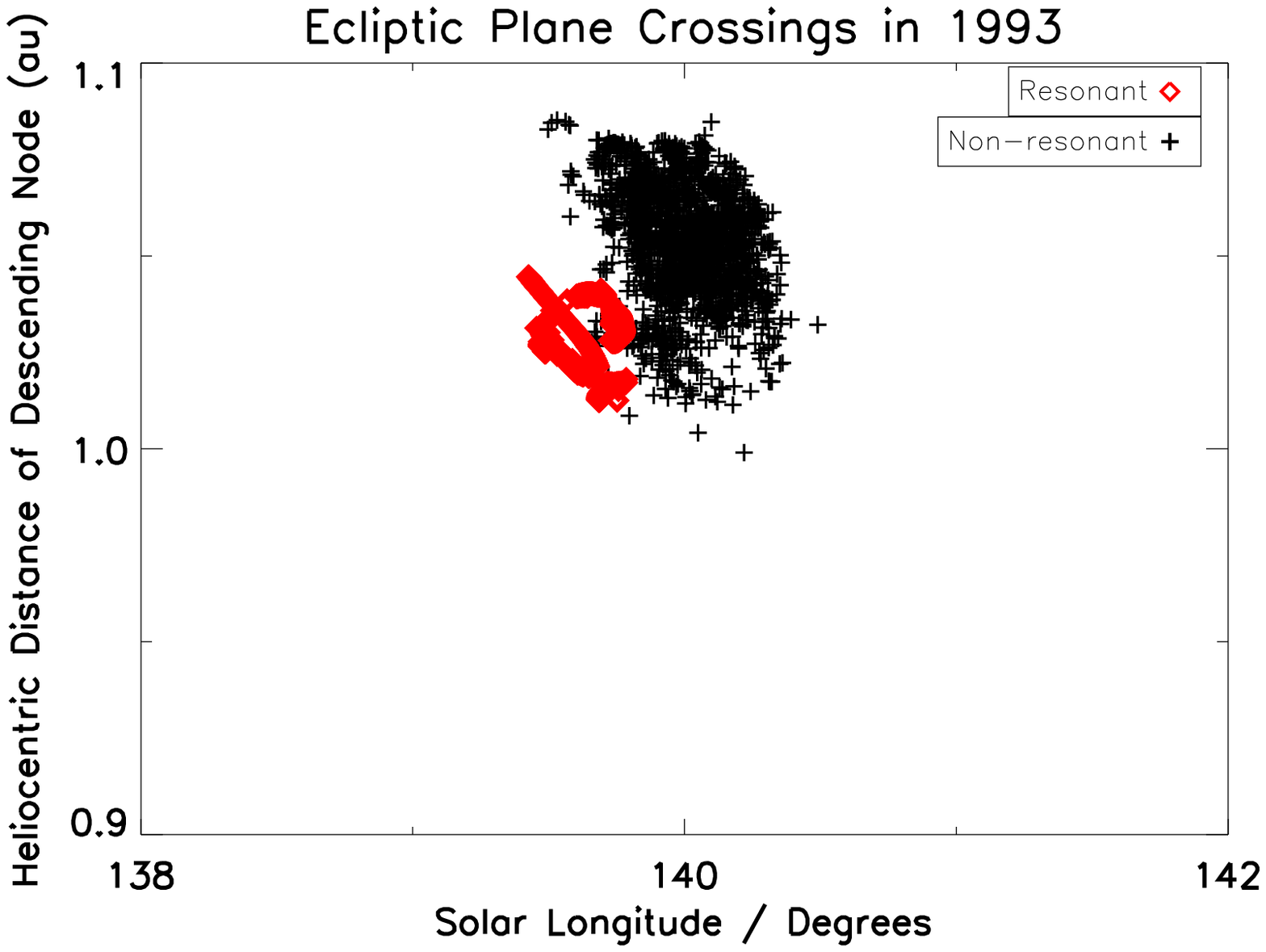}
\caption{Ecliptic plane crossings of Perseids:  heliocentric distance of
descending node versus solar longitude (J2000) in 1993 \AD, for particles
that are \js\ resonant and non-resonant (circulating in both
Jovian and Saturnian 2-body MMR and 3-body MMR) for particles with different initial ranges in (a) M=0-5\degr\  (for resonant case) and M=45-50\degr\  (for non-resonant case) (b) M=5-10\degr\  (resonant) and M=50-55\degr\  (non-resonant)  (c) M=350-355\degr\  (resonant) and M=260-265\degr\  (non-resonant) for subsequent verifications of resonant and non-resonant datasets from Figure \ref{aM} . Resonant cases show denser
structures whereas significant dispersion can be seen comparatively in the 
non-resonant meteoroids. Both resonant and non-resonant cases consist of same number (i.e.\ 2000 clones each) of particles. Integration start time = 69 \BC\ perihelion time of 
comet 109P.}  
\label{rdsol}
\end{figure}

In order to check whether 3-body MMR causes compact structures in the real
Perseid stream, we set up independent integrations to look at the evolution of multiple sets of resonant and non-resonant particles separately and verify the contrast in
their dynamics during present times. In the resonant case, 2000 particles
were integrated from 109P/Swift-Tuttle's 69 \BC\ return ($M$=0-5, 5-10, 350-355, 355-360\degr\ at JD 1696460.0;
offset chosen by looking at Fig.\ \ref{aM}). Initial
$a$ varied from 24.434 to 24.634 in steps of $ 5 \times 10^{-5}$ au while
initial $q$, $i$, $\omega$ and $\Omega$ are kept identical to the parent
body. All parameters were the same for the non-resonant particles except the
initial mean anomaly was adjusted appropriately (ranges in $M$=45-50, 50-55, 260-265, 265-270\degr\ at JD 1696460.0;
offset chosen by looking at Fig.\ \ref{aM}) so that the evolution of
multiple sets of non-resonant Perseids can be studied. We verified that no random close
encounters with planets significantly affecting the evolution occurred during
this offset time and the same number of particles each were always used in both the resonant and non-resonant cases. This is important if we are to compare densities in both evolutions on a like to like basis. 

Figure \ref{rdsol} shows the ecliptic-plane crossings for Perseids in
1993 \AD. The contrast in densities between resonant and non-resonant cases
(cf.\ comparisons 
of libration and circulation regimes in figure 3 of Emel'yanenko \& Bailey
1996) for the same number of particles (2000 clones each) can be clearly seen. This specifically shows the compactness of
resonant dust trails in a phase space which primarily deals with the spatial
spread near the Earth. Some Perseid particles are seen to directly intersect Earth 
in 1993 (nodal crossing times for resonant particles were verified to confirm intersecting possibilities
at this epoch). The active role of different Jovian and
Saturnian 2-body MMR in preserving compact dust trails for a long time has
previously been demonstrated (Asher et al.\ 1999; Rendtel 2007; Sato \&
Watanabe 2007; Sekhar \& Asher 2013, 2014). Like these 2-body MMR cases, we
find that \js\ MMR (3-body) can produce similar compact structures in the
Perseid stream and lead to enhanced meteor activity.

\begin{figure}
(a)\\[-\baselineskip]
\includegraphics[width=\columnwidth]{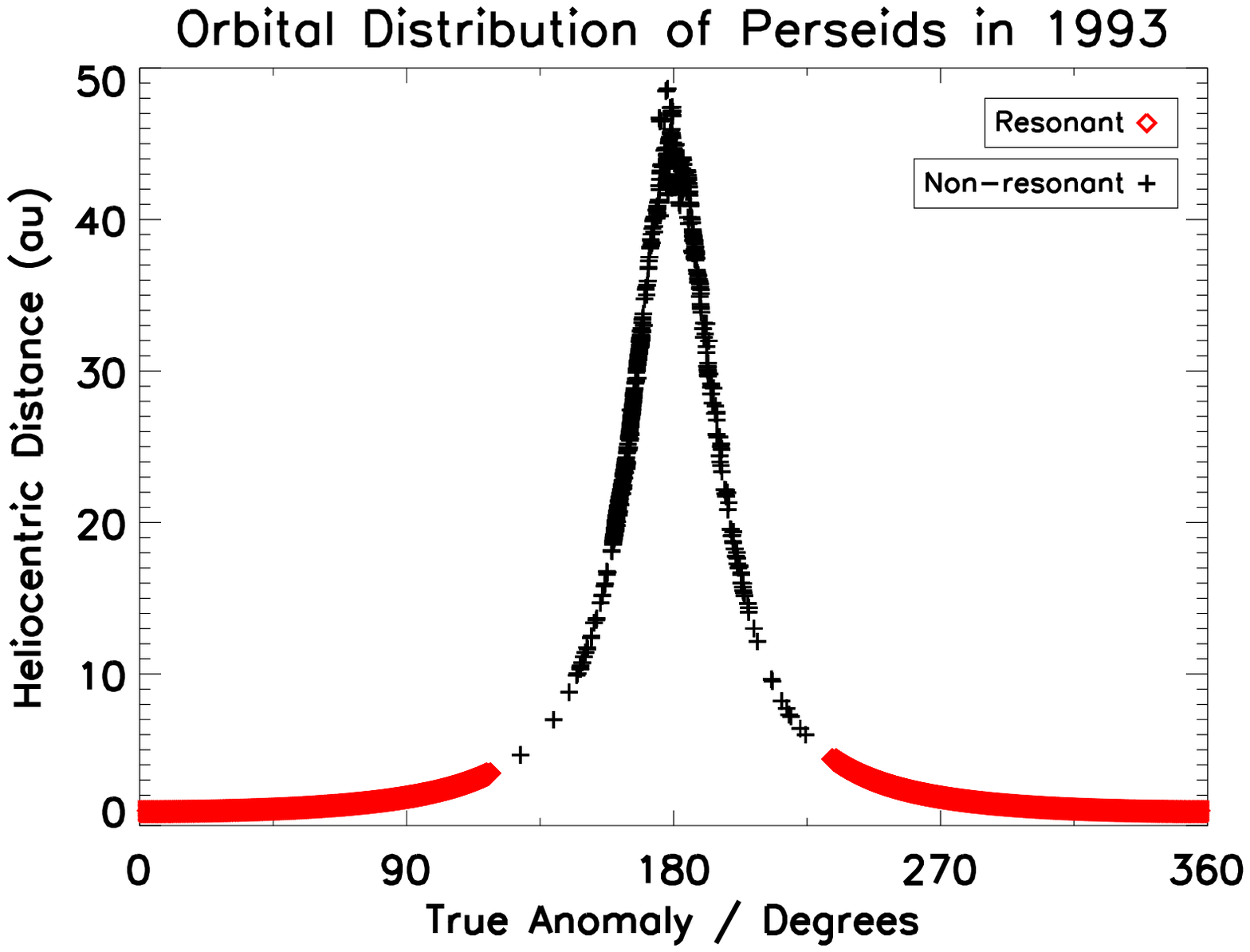}
\\[-3mm]
(b)\\[-\baselineskip]
\includegraphics[width=\columnwidth]{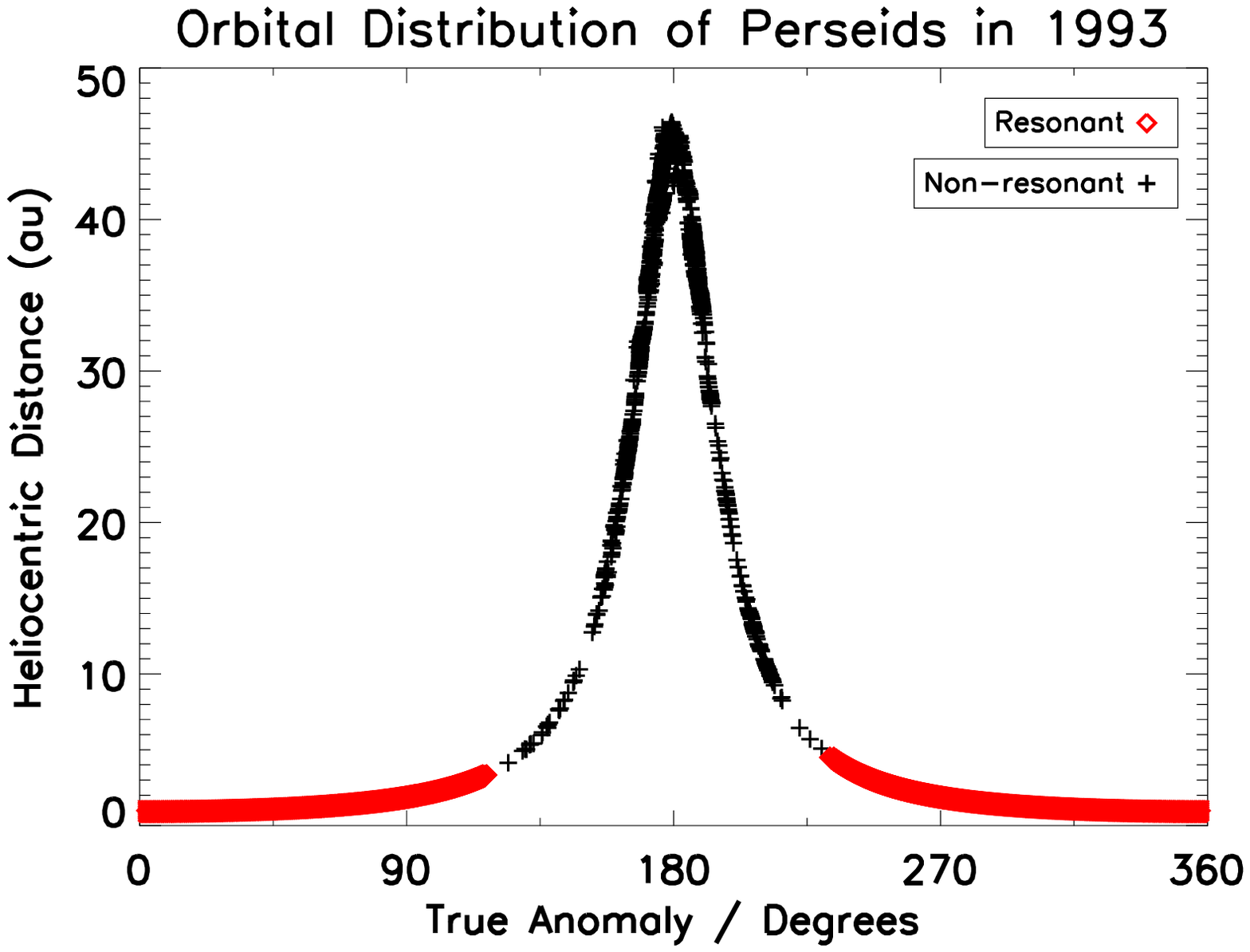}
\caption{Distribution along the entire
orbit -- i.e.\ heliocentric distance versus true anomaly -- of Perseid particles
for \js\ case (3-body MMR) and non-resonant Perseids (i.e.\ neither 2-body nor
3-body resonant) in 1993 \AD\ (for integrations starting in 69 \BC) for different initial ranges in (a) M=0-5\degr\  (for resonant case) and M=45-50\degr\  (for non-resonant case) (b) M=350-355\degr\  (resonant) and M=260-265\degr\  (non-resonant) for double checking the dynamical differences inferred from Figure \ref{aM}. Both resonant and non-resonant cases consist of same number (i.e.\ 2000 clones each) of particles. The resonant meteoroids show more compactness in the along-orbit dimension compared to the near 360\degr\ spread around the entire torus which can be seen in the case of non-resonant meteoroids. Moreover in 1993 a substantial population of the resonant meteoroids come close to the Earth and many are able to intersect Earth which can create enhanced activity.}  
\label{rf}
\end{figure}

\begin{figure}
(a)\\[-\baselineskip]
\includegraphics[width=\columnwidth]{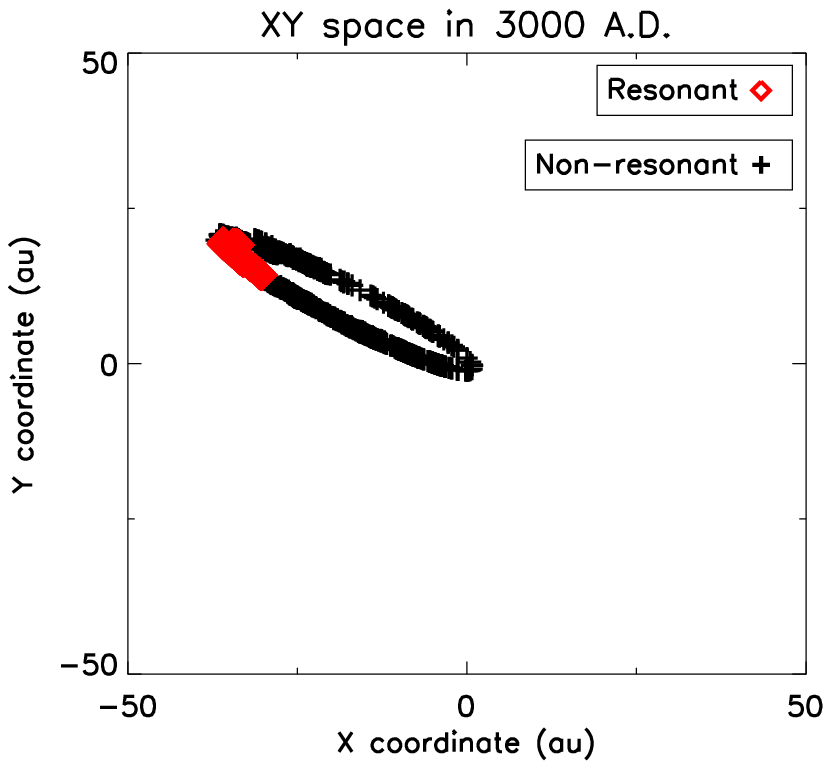}
\\[-3mm]
(b)\\[-\baselineskip]
\includegraphics[width=\columnwidth]{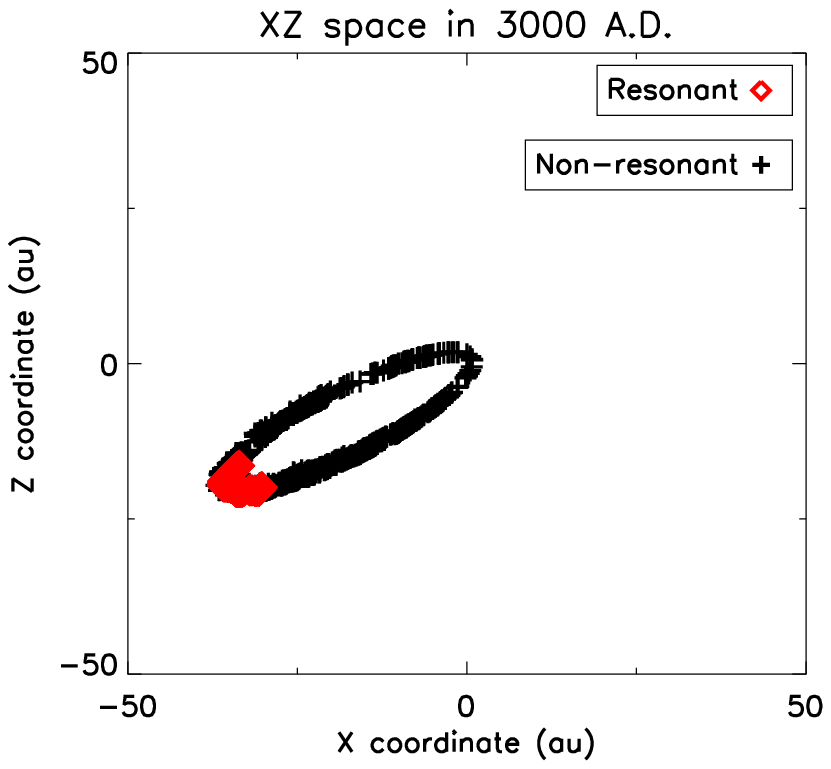}
\caption{(a) Ecliptic XY space (b) XZ space of Perseid particles for 2-1J+2S case (3-body MMR) and non-resonant Perseids (i.e.\ neither 2-body nor 3-body
resonant) in 3000 AD (for integrations starting in 69 BC). The plots show
that chaotic effects do not disrupt the stream geometry (even though chaos is deeply interconnected with 3-body resonances in general compared to 2-body resonances) and the resonant meteoroids remain confined to the toroidal structure thus preserving the typical meteoroid stream structure for more than 3 kyr. Both resonant and non-resonant cases consist of same number (i.e.\ 2000 clones each) of particles.}
\label{xyz}
\end{figure}

\begin{table}
\centering
\caption{Close-approach times and minimum close-approach distances between Earth and 3-body resonant meteoroids for 69 BC (2-1J+2S resonance) and 1862 AD (2+1J-3S resonance) ejections for different values of radiation pressure
$\beta$. Cutoff for maximum close-approach distance was set to 1 Hill radius of the Earth. 
}
\label{clotable}
\begin{tabular}{@{}rcccc@{}}
\hline
Ejection Epoch &$\beta$ & Close Approach   & Minimum Close  \\
                             &               &  Year                                        &  Approach Distance (au) \\ 
  &                                     &                               &      \\
\hline
69 BC return             & 0.0           & 1297 & 0.00045     \\
(2-1J+2S MMR)           &                    & 1460 & 0.00064 \\
                                       &                      & 1736 & 0.00036\\
                                      &                        & 1880 & 0.00068\\
                                       &                       & 1992 & 0.00013\\
                                       &                       &2111  & 0.00015\\
                                       \\
                                      & 0.001           & 1419 & 0.00056     \\
                                       &                    & 1534 & 0.00081 \\
                                      &                      & 1764 & 0.00098\\
                                       &                      & 1568& 0.00043\\
                                      &                        & 1993 & 0.00012\\
                                       &                    & 2108 & 0.00010\\
                                       \\
                                     & 0.01           & 943 &     0.0090     \\
                                       &                    & 1071 & 0.0091 \\
                                      &                      & 1535 & 0.0059\\
                                       &                      & 1654 & 0.0071\\
                                      &                        & 1882 & 0.0097\\
                                       &                        & 2069 & 0.0057\\
                                       \\
                                       & 0.1           & NA & Hyperbolic     \\
                                        \\                                

1862 AD return       & 0.0           & 1984 & 0.00030     \\
(2+1J-3S MMR)         &                    & 2215 & 0.00069 \\
                                      &                      &2330 & 0.00040\\
                                       &                      & 2446 & 0.00021\\
                                      &                        & 2562 & 0.00031\\
                                       &                        & 2678 & 0.00022\\
                                       \\
                                    & 0.001           & 1980 & 0.00032    \\
                                       &                    & 2211 & 0.00062 \\
                                      &                      & 2442 & 0.00063\\
                                       &                      &2558 & 0.00072\\
                                      &                        & 2674 & 0.00086\\
                                       &                        & 2790 & 0.00065\\
                                       \\
                                      & 0.01           & 2122 & 0.0071     \\
                                       &                    & 2254 & 0.0098 \\
                                      &                      & 2502 & 0.0063\\
                                       &                      &2633 & 0.0025\\
                                      &                        & 2729 & 0.0052\\
                                       &                        & 2870 & 0.0065\\
                                       \\
                                        & 0.1           & None & No approach    \\

\hline
\end{tabular}\\
\end{table}

The resonant particles can be seen to occupy only a small part of the entire
orbit (close to perihelion and the Earth's orbit) in contrast to
non-resonant particles spreading along the whole orbit in Figure
\ref{rf}(a) and (b) for multiple sets of particles from ($a$,$M$)
phase space. This shows how particles are packed in the along-orbit
dimension (in addition to spatial dimensions across ecliptic plane discussed
above), with the non-resonant meteoroids dispersed
over a large range of heliocentric distances while the resonant particles are concentrated over a small range in heliocentric distances including, at this epoch,
the orbit space near the Earth. Since the number of particles is the
same and the initial $a$ ranges are the same, between resonant and
non-resonant cases, the consistent, significant
difference in the distribution in both phase spaces (i.e.\ ecliptic
plane plus along-orbit direction) for multiple sets of resonant and
non-resonant particle pairs directly indicates the contrast between the
long-term evolution and structure of 3-body resonant
and non-resonant Perseid meteoroids. 

\begin{figure}
(a)\\[-\baselineskip]
\includegraphics[width=\columnwidth]{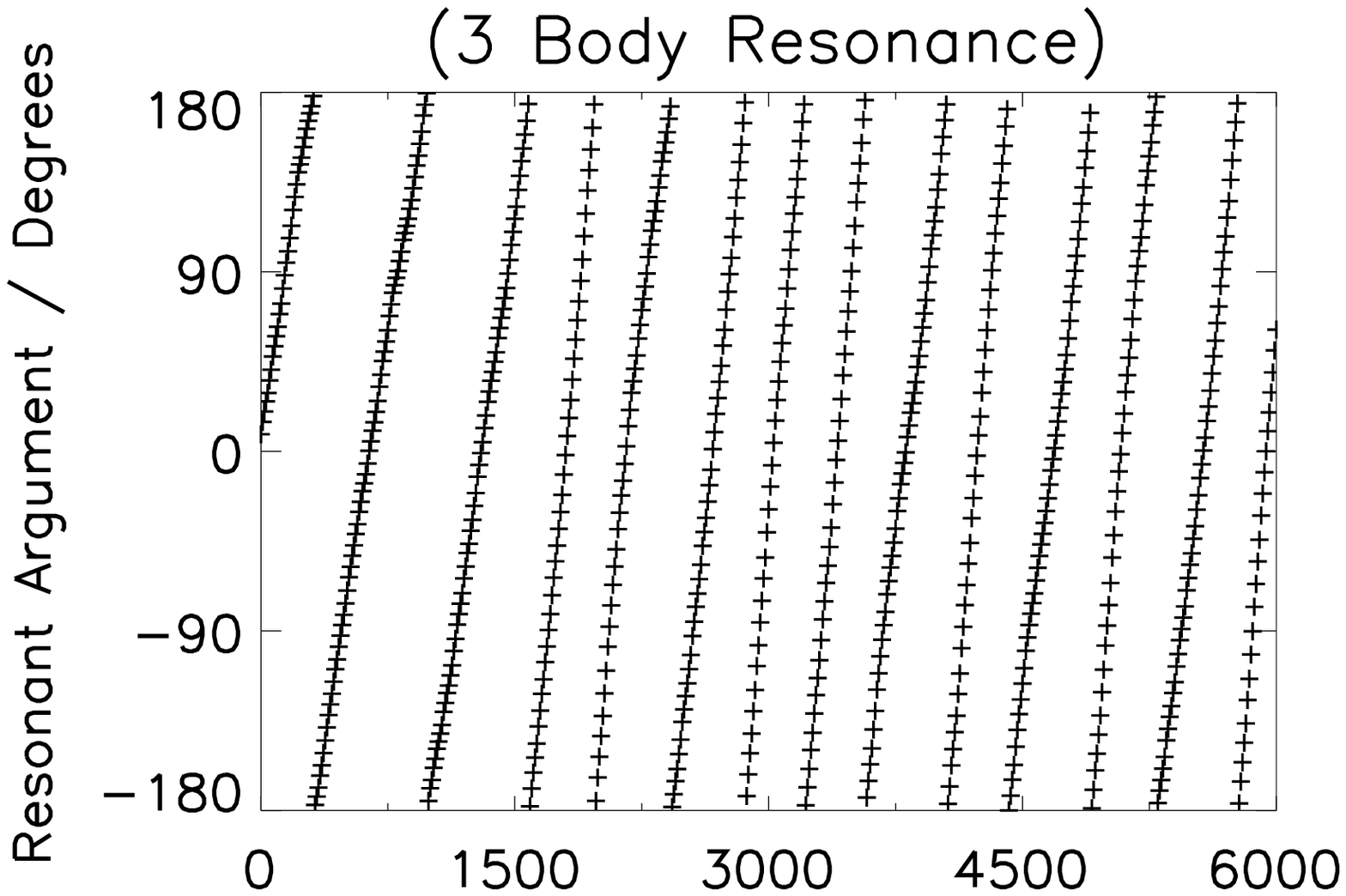}
\\[-5mm]
(b)\\[-\baselineskip]
\includegraphics[width=\columnwidth]{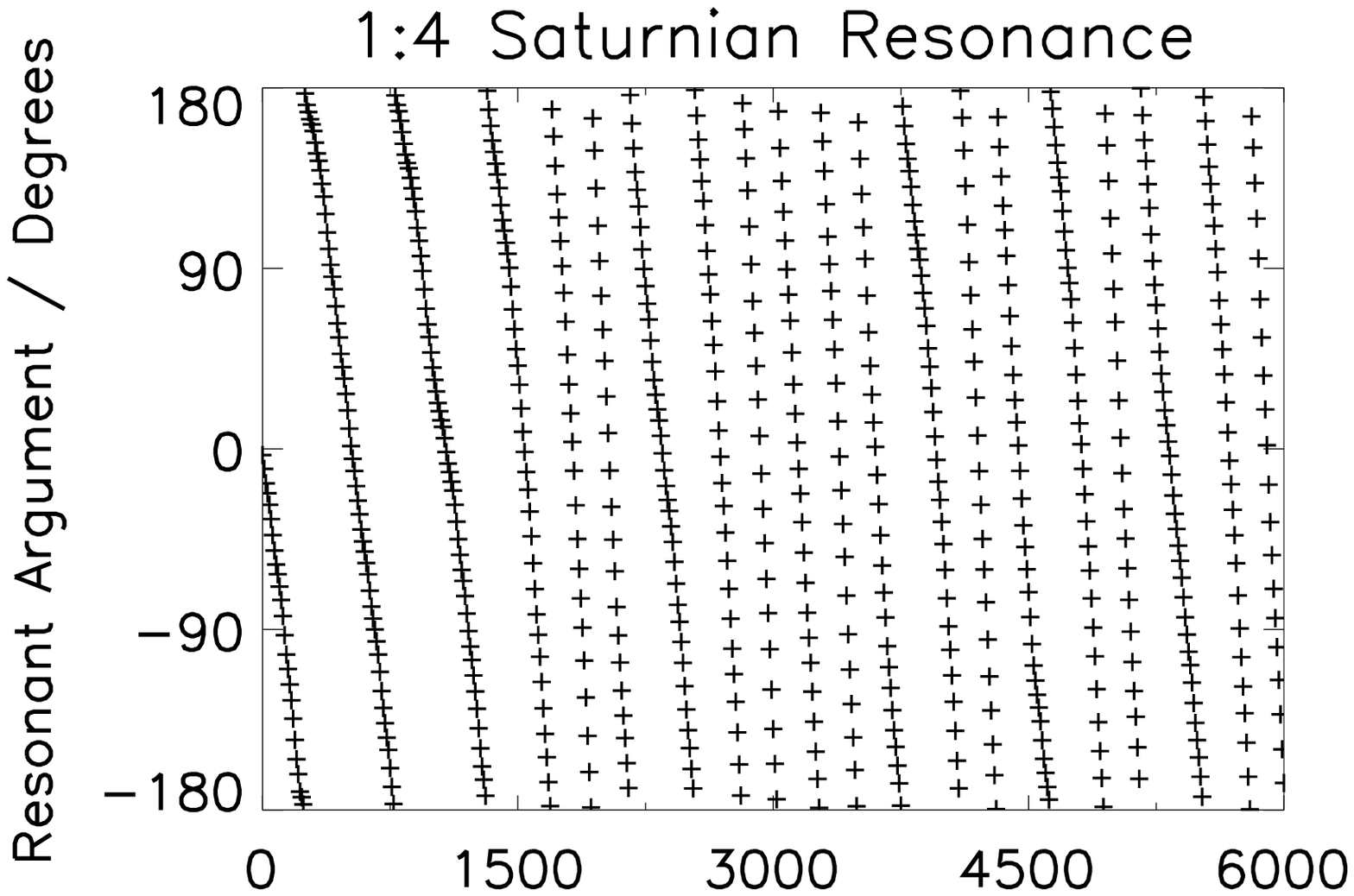}
\\[-5mm]
(c)\\[-\baselineskip]
\includegraphics[width=\columnwidth]{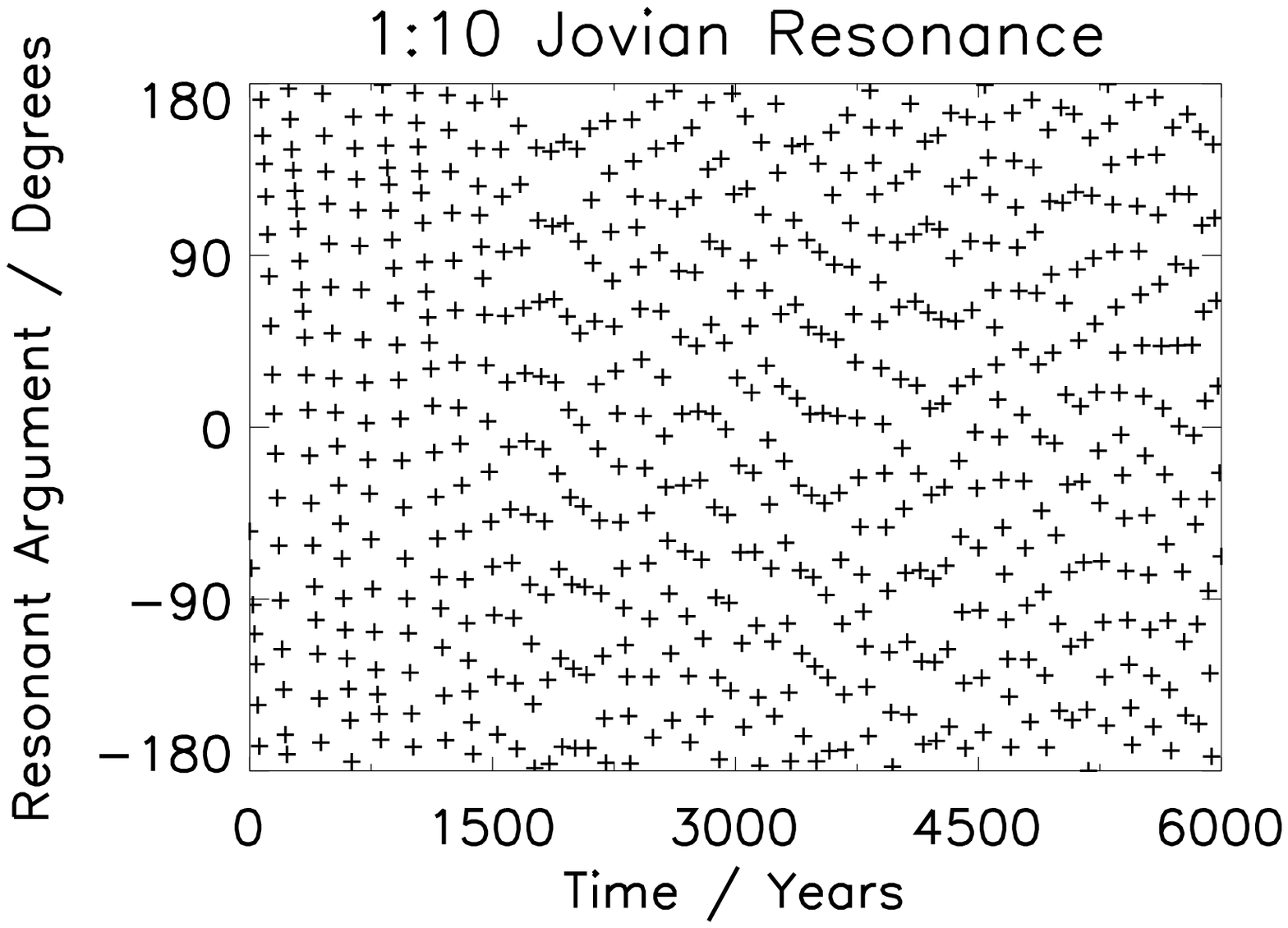}
\caption{(a) Circulation of \js\ resonant argument for a Perseid
test particle (initial $a$=24.41 au), confirming absence of 3-body MMR
involving Jupiter and Saturn simultaneously.  Outer circulation of (b) 1:10
Jovian (2-body MMR) and (c) 1:4 Saturnian (2-body MMR) resonant arguments for
the same particle during same time frame confirm absence of 1:10 Jovian and 1:4 Saturnian separately. Starting epoch (zero time) is JD 1789915.96 = 188 \AD\ return.} 
\label{PERsigma188}
\end{figure}

\begin{figure}
(a)\\[-\baselineskip]
\includegraphics[width=\columnwidth]{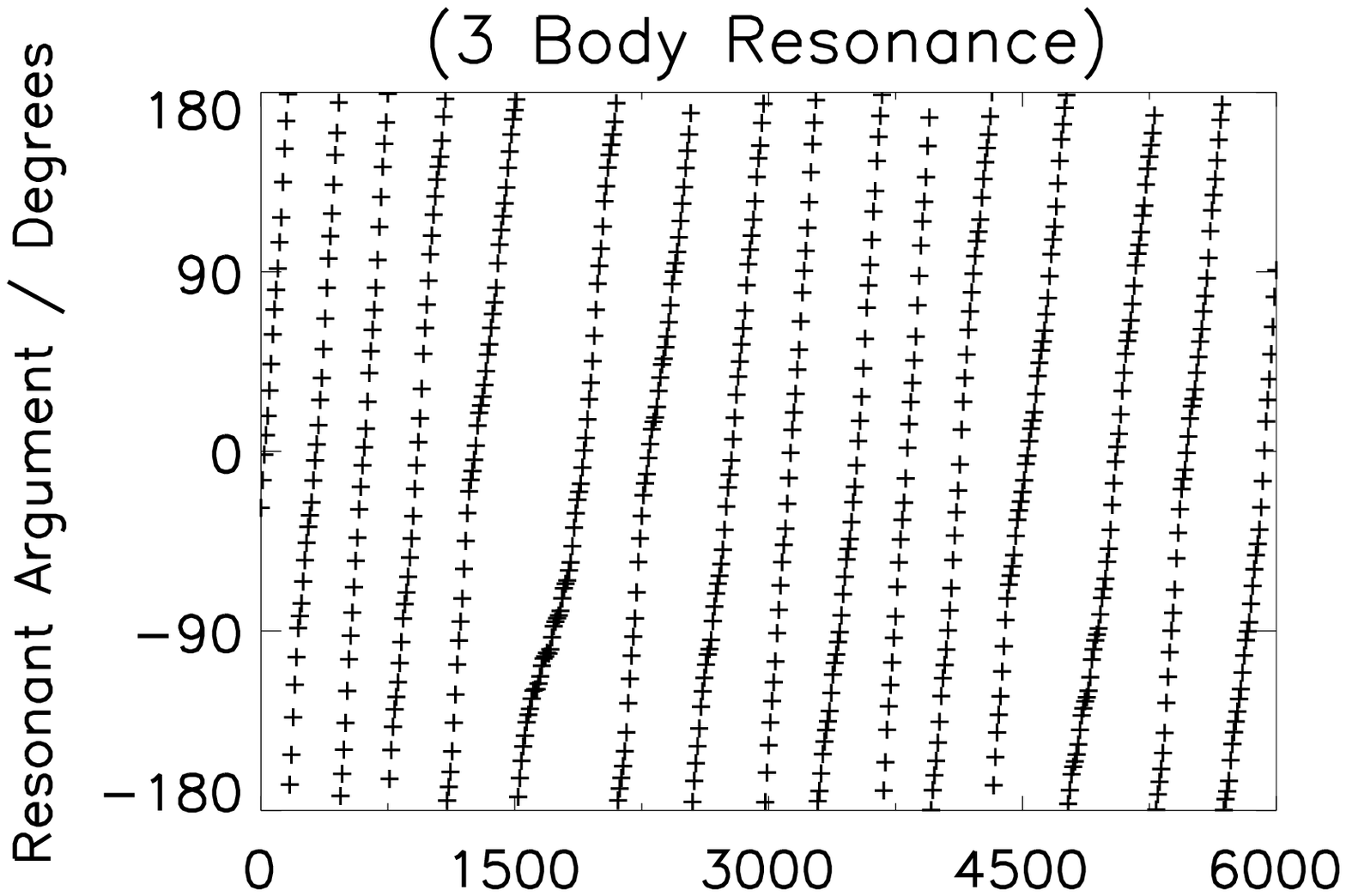}
\\[-5mm]
(b)\\[-\baselineskip]
\includegraphics[width=\columnwidth]{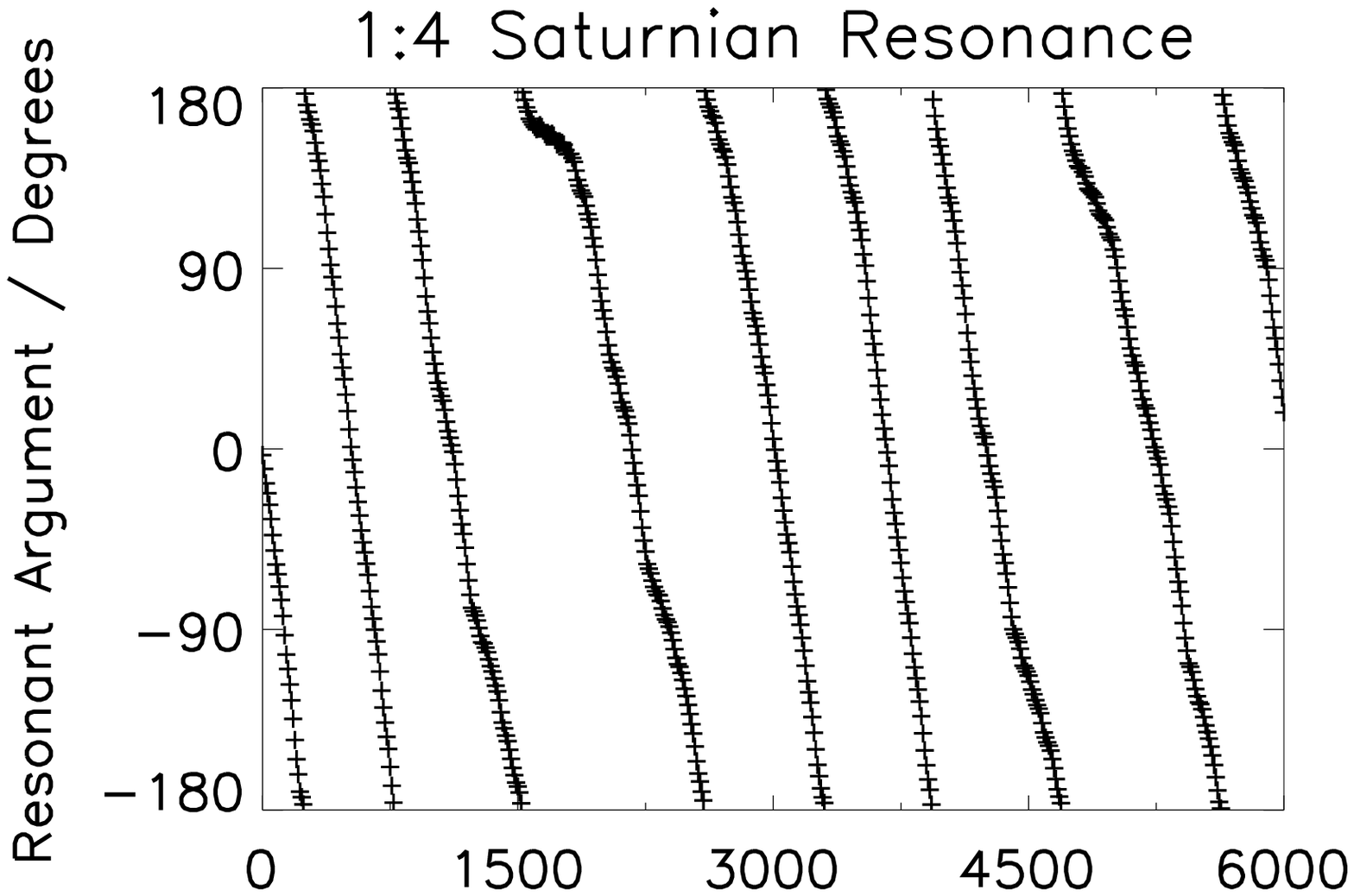}
\\[-5mm]
(c)\\[-\baselineskip]
\includegraphics[width=\columnwidth]{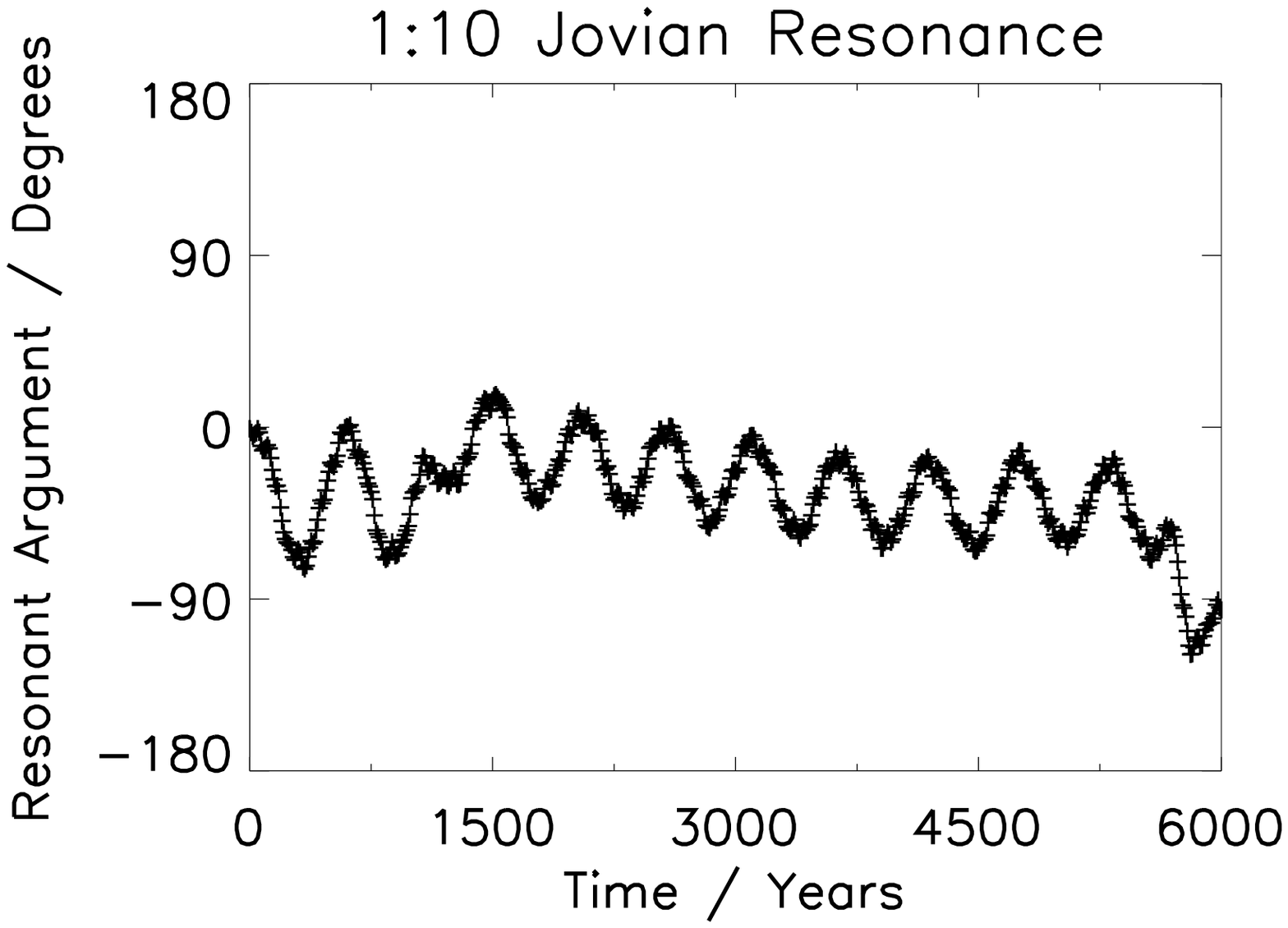}
\caption{Circulation of (a) 2-1J+2S resonant argument
confirming absence of 3-body MMR
involving Jupiter and Saturn simultaneously and (b) 1:4 Saturnian (2-body MMR) resonant argument showing absence of 1:4 Saturnian for a Perseid
test particle (initial $a$=24.41 au). Small-amplitude libration of (c) 1:10  
Jovian (2-body MMR) for the same particle during same time frame confirms presence of strong 1:10 Jovian.
Starting epoch (zero time) is JD 2355652.3  = 1737 \AD\ return.} 
\label{PERsigma1737}
\end{figure}

\begin{figure}
(a)\\[-\baselineskip]
\includegraphics[width=\columnwidth]{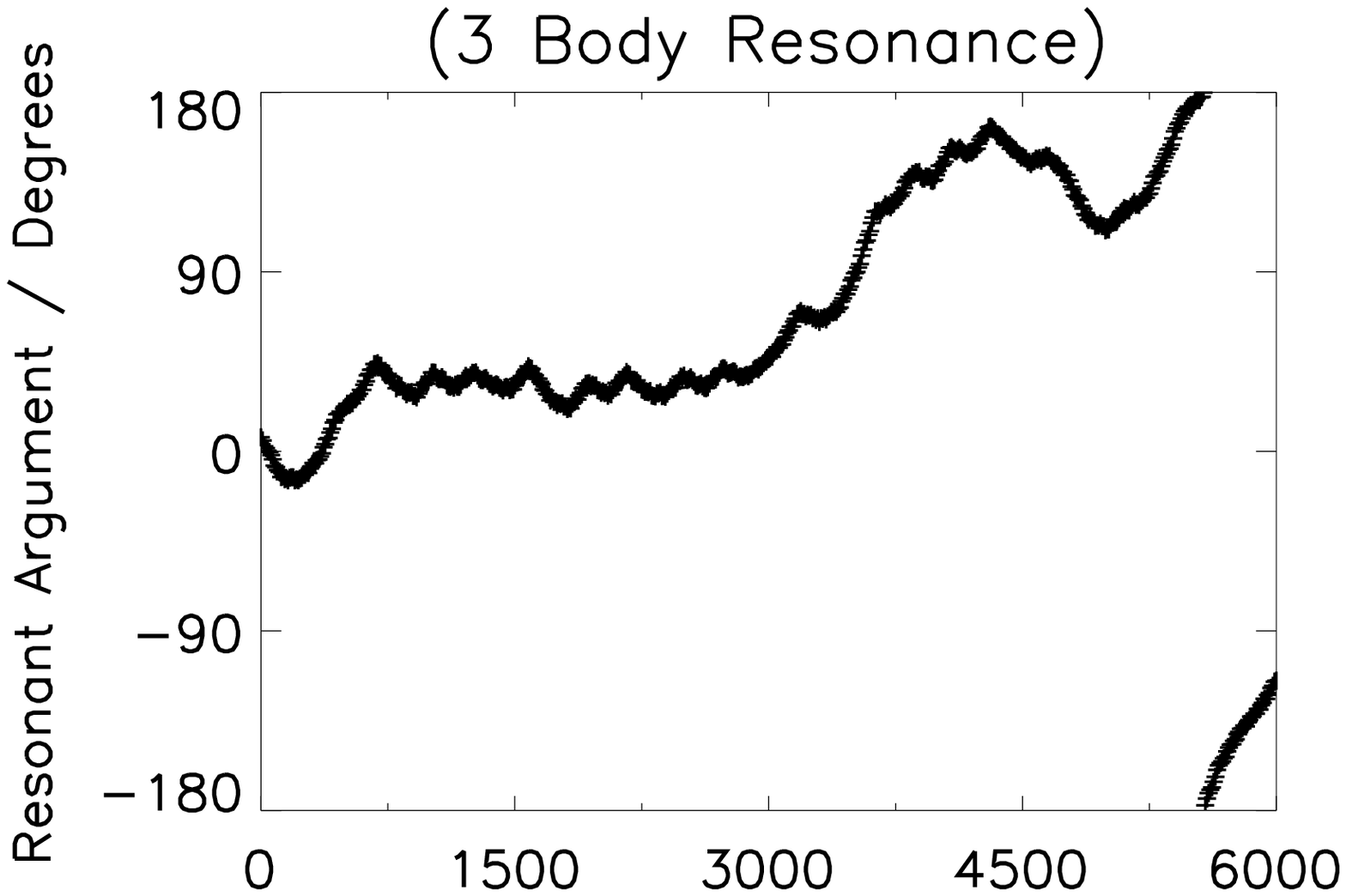}
\\[-5mm]
(b)\\[-\baselineskip]
\includegraphics[width=\columnwidth]{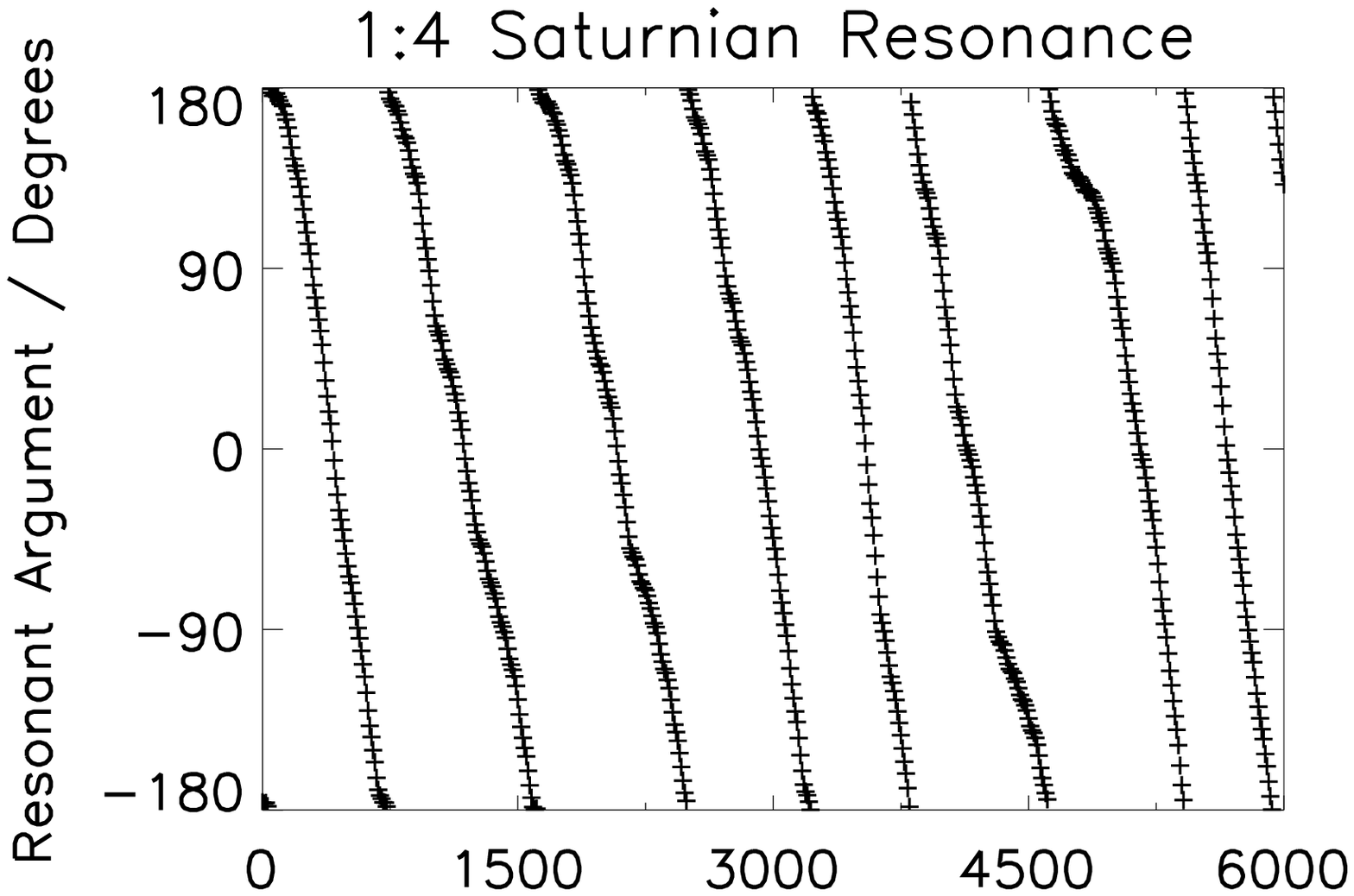}
\\[-5mm]
(c)\\[-\baselineskip]
\includegraphics[width=\columnwidth]{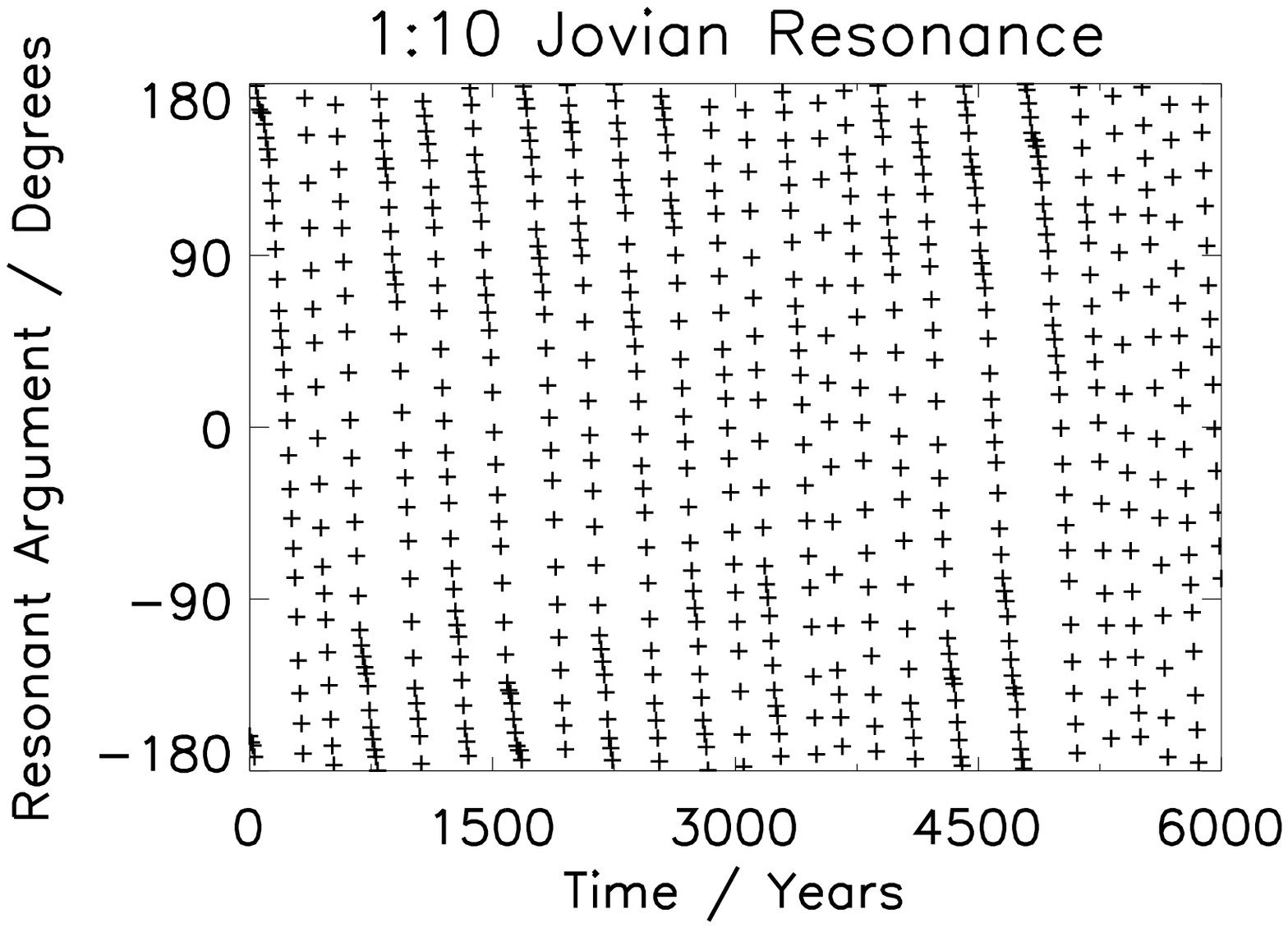}
\caption{(a) Libration of 2+1J-3S resonant argument for a Perseid 
test particle (initial $a$=23.40 au), confirming presence of 3-body MMR 
involving Jupiter and Saturn simultaneously.  Outer circulation of (b) 1:10
Jovian (2-body MMR) and (c) 1:4 Saturnian (2-body MMR) resonant arguments for
the same particle during same time frame confirm absence of 1:10 Jovian
and 1:4 Saturnian separately. Starting epoch (zero time) is JD 2401375.9 = 1862 \AD\ return.}  
\label{PERsigma1862}
\end{figure}

\begin{figure}
(a)\\[-\baselineskip]
\includegraphics[width=\columnwidth]{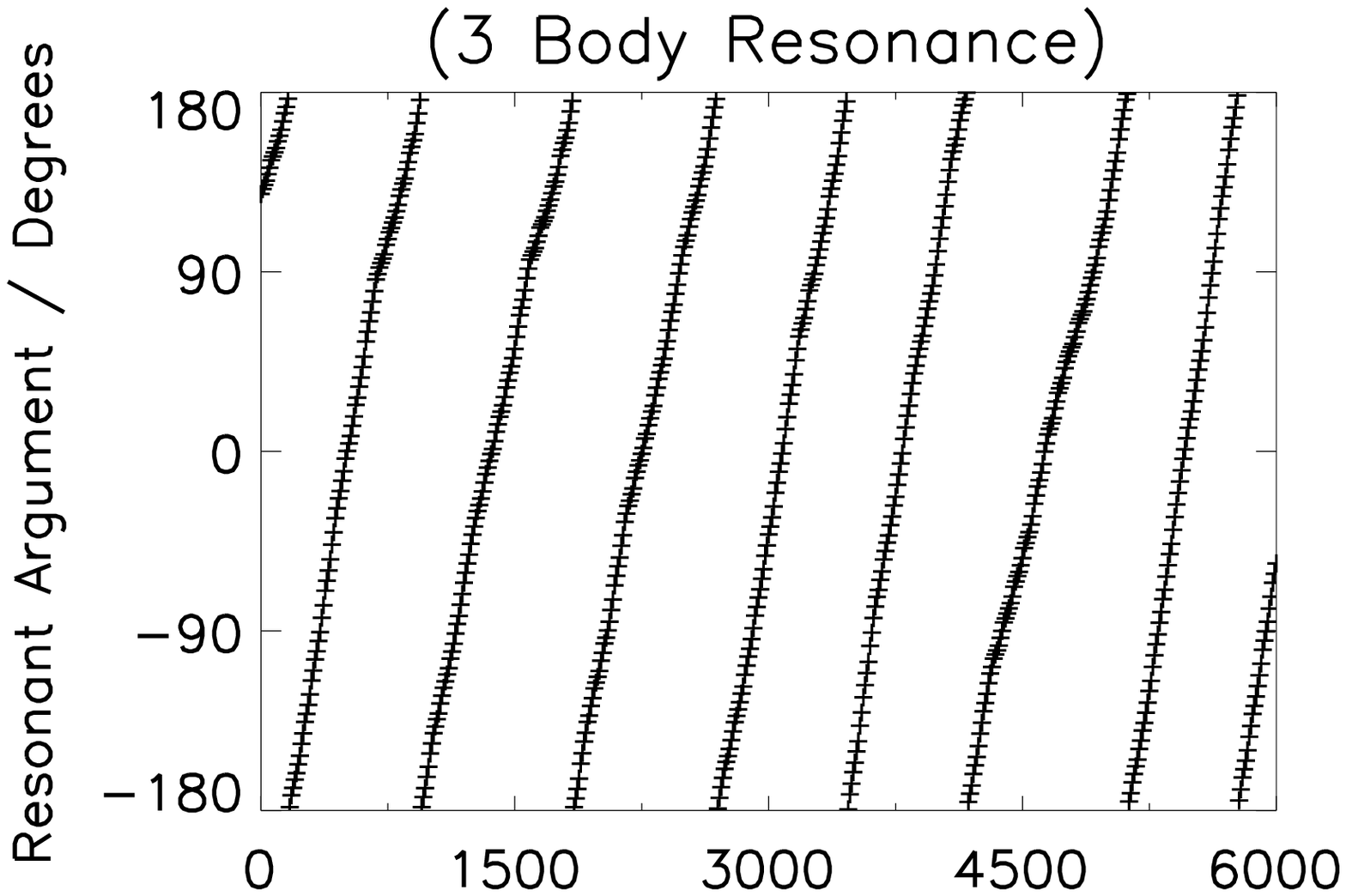}
\\[-5mm]
(b)\\[-\baselineskip]
\includegraphics[width=\columnwidth]{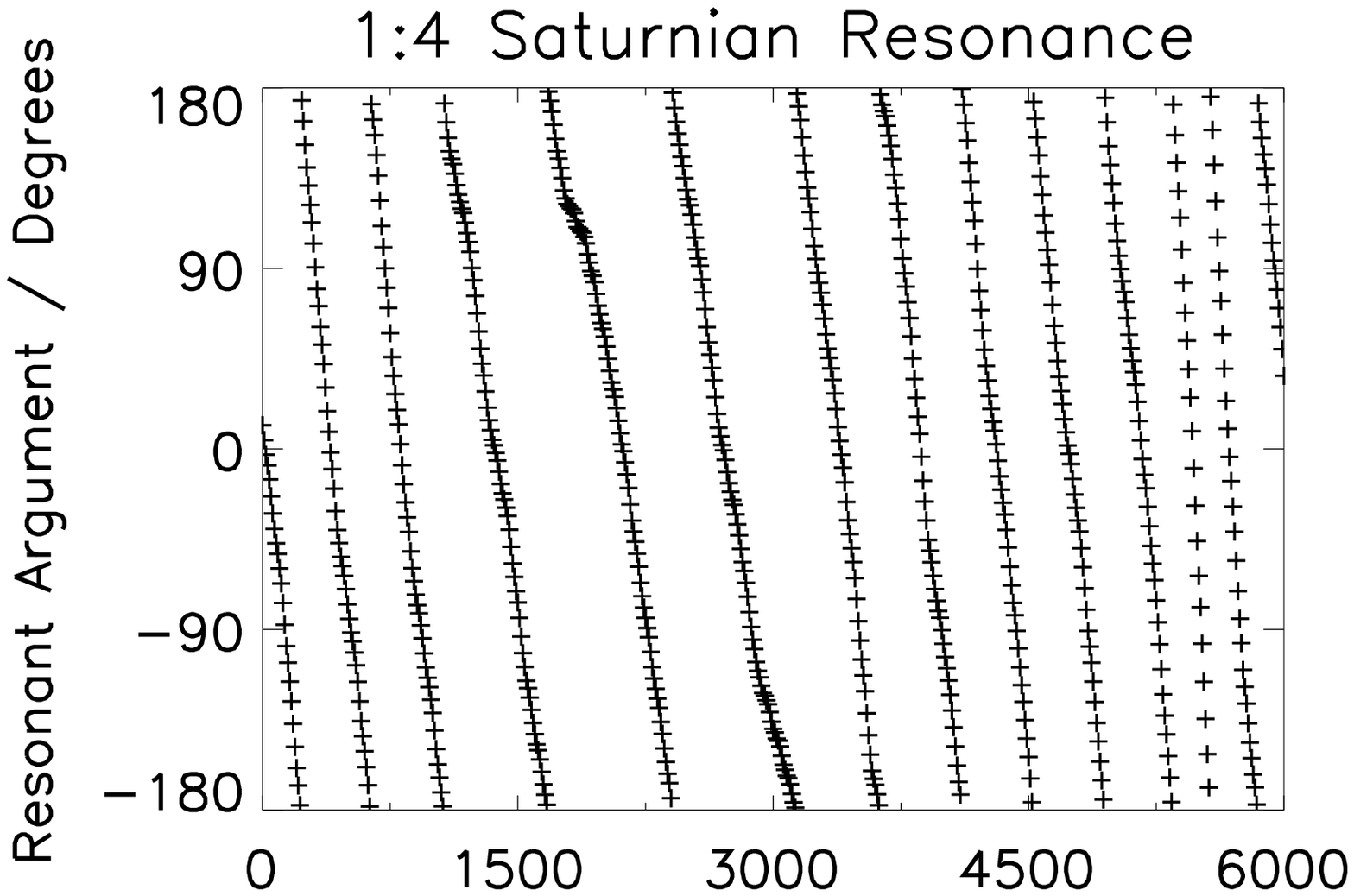}
\\[-5mm]
(c)\\[-\baselineskip]
\includegraphics[width=\columnwidth]{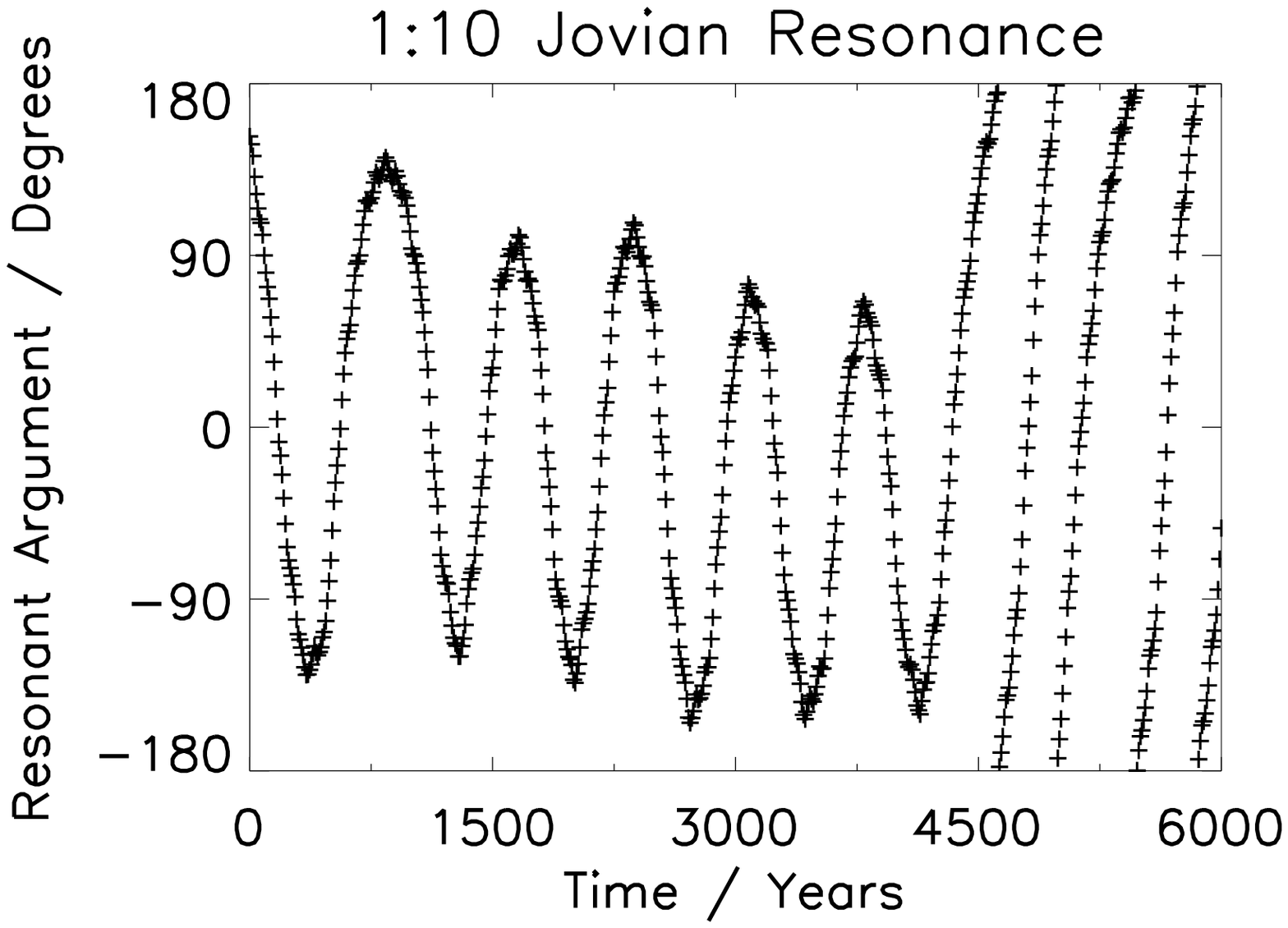}
\caption{Circulation of (a) 2-1J+2S resonant argument confirming absence of 3-body MMR
involving Jupiter and Saturn simultaneously and (b) 1:4 Saturnian (2-body MMR) resonant argument showing absence of 1:4 Saturnian for a Perseid
test particle (initial $a$=24.41 au), Libration of (c) 1:10
Jovian (2-body MMR) for the same particle during same time frame confirms presence of 1:10 Jovian.
Starting epoch (zero time) is JD 2448968.5 = 1992 \AD\ return, the
most recent observed return of 109P/Swift-Tuttle (Marsden \& Williams 2008).}  
\label{PERsigma1992}
\end{figure}

Heliocentric distance $r$ (Fig.\ \ref{rf}) does not directly give the
exact positions in cartesian space -- i.e.\ the $r$ value can in principle come
from any possible combinations of XYZ coordinates satisfying the Pythagorean
expression. Hence it is instructive to double check by explicitly looking at
their distribution in XYZ space to ensure all resonant particles remain inside the stream structure. Figures \ref{xyz}(a) and (b) show the XY and
XZ space of resonant and non-resonant particles in 3000 \AD\
indicating that the resonant particles stay very much as
part of the general Perseid stream orbit into the distant future (for the purpose of meteoroid stream predictions) and no random chaotic effects occur
which make them deviate from the typical toroidal meteoroid stream structure.
Often a longer time of many 10 kyr or more is required for chaotic
effects to dominate in orbital evolution of most small bodies.
But because 3-body MMR is more closely linked with small stability islands
closely surrounded by areas of severe chaotic nature (see figure 9 in
Beaug\'e et al. 2008), Figure \ref{xyz} (a) and (b) serves as a simple
verification to rule out convincingly any random chaotic affects unsettling
particles from a typical meteoroid stream
torus structure. Ruling out short-term chaos in particle orbits is usually a
necessary (but not sufficient) condition for stable resonance. 

A systematic check was done to look for the general possibilities of 3-body resonant particles coming near the Earth in the recent past and future from all the observed perihelion passages of the parent comet. 100 clones each were ejected at all the observed perihelion passages of the comet
(69 \BC, 188 \AD, 1737 \AD, 1862 \AD\ and 1992 \AD\ returns) around each of
the six nominal 3-body resonant locations in Table 1. Only the 69 \BC\ and 1862 returns are favourable for trapping Perseid
particles in strong 3-body MMRs: representative particles are plotted in
Figures \ref{PERsigma}, \ref{PERsigma188}, \ref{PERsigma1737}, \ref{PERsigma1862}
and \ref{PERsigma1992}.
And out of all the 3-body MMRs in Table 1, the 69 \BC\ return is favourable for both \js\ and 2+1J-3S configurations. It is found that 1862 \AD\ is favourable for just 2+1J-3S (representative particle in figure \ref{PERsigma1862}). The other resonant arguments (for configurations listed in Table 1) than the \js\ and 2+1J-3S do not show libration for starting times corresponding to any of these perihelion passages. During the remaining observed perihelion returns (188, 1737, 1992), the comet's position is such that either it is mostly efficient to populate 1:10 Jovian MMR (which is the case in 1737 and 1992; representative particles in figures \ref{PERsigma1737} and \ref{PERsigma1992}) or it populates none of the 2-body or strong 3-body MMRs (representative particle in figure \ref{PERsigma188}) discussed in this work. 

Thus out of the known observed perihelion returns of the comet, those of
69 \BC\ and 1862 are more favourable for inducing either or both of two
distinct 3-body MMRs. Hence we integrate two distinct 3-body resonant cases
(from the two respective perihelion passages) consisting of 2000 particles
each (covering the whole $a$ range favouring both these 3-body MMR
configurations) for 3 kyr forward in time under the influence of four
different values of the radiation pressure parameter $\beta$ = 0, 0.001, 0.01
and 0.1 (corresponding to different particle sizes). In total therefore, 8
separate integration sets were done to look for possible Earth-meteoroid
intersection possibilities. 

Close encounters between 3-body resonant particles and Earth were tracked using MERCURY for clones ejected from these two perihelion returns,
namely the 69 \BC\ (favouring 2-1J+2S and 2+1J-3S resonance configurations) and 1862 \AD\
(favouring 2+1J-3S) returns.
Table \ref{clotable} shows the close-approach distances between the 3-body resonant particles and the Earth for past and future
years for different values of $\beta$. Sufficient tests (using critical
angle and trajectory tests) were done to ensure that the particles listed in
Table \ref{clotable} exhibit 3-body resonance (after the incorporation of
radiation pressure) and do not show the Jovian and Saturnian 2-body
resonances during its evolution. The values in the
Table show that 3-body resonant particles come near the
Earth in the past and future time frames thereby indicating the possibilities
for enhanced meteor activity from these 3-body resonant dust trails. Our
integrations show many 3-body resonant meteoroids intersecting the Earth in
1992 and 1993. Interestingly Jenniskens et al. (1998) has reported enhanced
Perseid activity between 1989--1996 with peak meteor activity in 1993. It is
possible that 3-body resonant meteoroids also contributed to the enhanced
activity during this period (in addition to 2-body Jovian resonances which
have been correlated by Jenniskens et al. (1998) previously).

These cases mentioned in Table \ref{clotable} pertain to 3-body resonant particles in a wide range of semi-major axis (covering the entire resonant-zone span) with $\beta$ values of 0, 0.001, 0.01 and 0.1 ejected tangentially at perihelion, thus giving a general estimate of the nature of their evolution and fate. For extremely accurate meteor density or flux estimations/predictions at precise times, one needs to take multiple sets of particles with finer resolution in semi-major axis, extend the ejection arc (to pre- and post-perihelion points) and include more combinations of $\beta$ and ejection epochs. This is beyond the scope of this paper but an independent elaborate study is planned
for the future. We expect such a detailed study would help us to correlate
the specific observed Perseid features between 1989--1996 with specific
aspects of 3-body resonant simulations in an accurate way and come up with
more accurate future predictions.

\vspace*{-5mm}

\section{Conclusion and Future Work} 

We have shown that Perseid particles exhibit a unique 3-body resonance 
mechanism (formal nomenclature of these
MMR being \js \ and 2+1J-3S) close to 1:4:10 MMR involving
Perseids, Saturn and Jupiter respectively. Individual
particles can get trapped in these resonances for 
up to $\sim$4 kyr. Due diligence
was followed to ensure that none of these particles show 2-body MMR with
Jupiter or Saturn separately during the same time frame. Hence it is
confirmed that the resonant phenomenon is due to a 3-body mechanism rather
than the conventional 2-body mechanism. Furthermore it is seen that the dust 
trails can retain very compact structures typically for the order of a few
kyr and these compact dust trails can come near the Earth and intersect Earth in past and future times. In summary, this is the first theoretical example of a stable 3-body resonance in the realm of meteoroid streams.  

We found examples of 3-body resonant dust trails intersecting the Earth
in 1992 and 1993 which agrees with already observed enhanced Perseid activity
between 1989--1996 (as reported by Jenniskens et al.\ 1998). Further work
needs to be done to correlate the exact observed features with simulations
and give accurate predictions for the future. The future challenge is to run
detailed simulations, using finer resolution in the $a$ of ejected particles
and extending the ejection arc to pre- and post-perihelion intervals, in
order to find precise meteoroid-Earth intersection times and estimate the
density or flux of peak meteor activity and the duration of possible enhanced
activity due to these 3-body MMRs from the already identified favourable
perihelion passages. 

Because the 3-body resonance mechanism also favours clustering and increase
of density of meteoroids in space just like in the case of 2-body resonances,
one can expect spectacular meteor outbursts and storms on Earth from this
resonance mechanism in the future. Any prediction for such
a future outburst or storm in the near future will be of great theoretical as
well as observational value. Further work is planned in this direction so
that professional and amateur observers can be alerted and scientific
observing campaigns can be launched soon. 

\vspace*{-5mm}

\section*{Acknowledgments}

Sekhar acknowledges the Crater Clock project (235058/F20) based at CEED funded through the Centres of Excellence scheme (project number 223272) by the Research Council of Norway and thanks the support from project PI Dr Stephanie Werner. Sekhar and Asher thank the Department of Culture, Arts and Leisure of
Northern Ireland for funding astronomical research at Armagh Observatory.  
We are grateful to Tabar\'e Gallardo and an anonymous reviewer for
thorough and helpful comments and acknowledge the valuable contribution of Dr
Gallardo to the research community in making available his software to
compute locations and strengths of 3-body resonances.

\end{document}